\documentclass[journal,10pt,doublespace,doublecolumn]{IEEEtran}
\usepackage{amsmath}
\usepackage{graphicx,psfrag,epsf}
\usepackage{enumerate}
\usepackage{url} % not crucial - just used below for the URL 
\usepackage{hyperref}
\usepackage{url} 
\usepackage{amssymb,amsfonts,mathtools}
\usepackage{bm}
\usepackage{lipsum}
\usepackage{color, colortbl}
\usepackage[shortlabels]{enumitem}
\usepackage{etoolbox}
\usepackage{cite}
\usepackage[font=footnotesize]{caption}
\usepackage{cases}
\usepackage{booktabs}
\usepackage{xcolor}

\usepackage{algorithm}
\usepackage{balance}
\usepackage[noend]{algpseudocode}
\usepackage{subfig}
\usepackage{romannum}
\usepackage{comment}
\usepackage{dsfont}
\usepackage{multirow}

\newtheorem{definition}{Definition}
\newtheorem{remark}{Remark}
\newtheorem{theorem}{Theorem}

\newtheorem{lemma}{Lemma} %\newtheorem{lemma}[theorem]{Lemma}
\newtheorem{corollary}{Corollary}
\newtheorem{proposition}{Proposition}

\newtheorem{assumption}{Assumption}
\newtheorem{proof}{Proof}

\makeatletter
\patchcmd{\@makecaption}
  {\scshape}
  {}
  {}
  {}
\makeatletter
\patchcmd{\@makecaption}
  {\\}
  {.\ }
  {}
  {}
\makeatother

\DeclareMathOperator*{\argmin}{arg\,min}

\newcommand{\rom}[1]{\uppercase\expandafter{\romannumeral #1\relax}}
\newcommand\independent{\protect\mathpalette{\protect\independenT}{\perp}}
\def\independenT#1#2{\mathrel{\rlap{$#1#2$}\mkern2mu{#1#2}}}

\DeclareMathOperator{\sinc}{sinc}
\DeclareMathOperator{\tri}{tri}

\usepackage{xspace}
\usepackage{bbm}
\usepackage{mathrsfs}
\def\Var{\mathop{\rm Var}\nolimits}%
\def\Cov{\mathop{\rm Cov}\nolimits}%
\newcommand{\Fc}{\mathcal{F}}

\def\a{\alpha}

\def\d{\delta}

\def\s{\sigma}

\DeclareMathOperator\E{\sf E}

\newcommand{\U}{\mathrm{Unif}}

\def\textiid{i.i.d.\@\xspace}
\newcommand\iid{\ifmmode\text{ i.i.d. } \else \textiid \fi}

\newcommand{\Real}{\mathbb{R}}
\newcommand{\Integer}{\mathbb{Z}}

\allowdisplaybreaks

\begin{document}

\title{Causal Inference from Slowly Varying Nonstationary Processes}

\author{Kang~Du and Yu~Xiang% <-this % stops a space
\thanks{
The material in this manuscript was presented in part at the IEEE International Symposium on Information Theory 2021~\cite{Du2021}. Kang Du and Yu Xiang are with the Electrical and Computer Engineering Department, University of Utah, Salt Lake City, 
UT, 84112, USA (e-mail: kang.du@utah.edu, yu.xiang@utah.edu).  
}% <-this % stops a space

}

%\markboth{IEEE Transactions on Signal Processing}
%{Shell \MakeLowercase{\textit{et al.}}: IEEE Transactions on Signal Processing}

\maketitle

\pagenumbering{arabic}

\begin{abstract}
 
Causal inference from observational data following the restricted structural causal model (SCM) framework hinges largely on the asymmetry between cause and effect from the data generating mechanisms, such as non-Gaussianity or non-linearity. This methodology can be adapted to stationary time series, yet inferring causal relationships from nonstationary time series remains a challenging task. In this work, we propose a new class of restricted SCM, via a time-varying filter and stationary noise, and exploit the asymmetry from nonstationarity for causal identification in both bivariate and network settings. We propose efficient procedures by leveraging powerful estimates of the bivariate evolutionary spectra for slowly varying processes. Various synthetic and real datasets that involve high-order and non-smooth filters are evaluated to demonstrate the effectiveness of our proposed methodology.
\end{abstract}

\IEEEpeerreviewmaketitle

%\section{Introduction}

\section{Introduction} \label{sec:intro}
\IEEEPARstart{I}{nferring} causal relationships from observational data has drawn much attention in recent  years~\cite{shimizu2006linear, hoyer2009nonlinear,zhang2009post,peters2014identifiability}, following the pioneering works on structural causal models (SCMs) by Pearl~\cite{pearl2000models}. The main theoretical challenge lies in the identifiability of the causal structure, which is not possible for general SCMs. As a result, various classes of \emph{restricted} SCMs have been proposed including the linear non-Gaussian acyclic model (LiNGAM)~\cite{shimizu2006linear}, the non-linear additive noise models (ANMs)~\cite{hoyer2009nonlinear,peters2014causal}, and the post-nonlinear causal model~\cite{zhang2009post}. The structure identifiability can be proved either exactly~\cite{shimizu2006linear} or \emph{in generic cases}~\cite{hoyer2009nonlinear,peters2014causal,zhang2009post}, and  
 the key to this is to break the symmetry between cause and effect via structural assumptions such as non-Gaussianity or non-linearity.

%Specifically, consider a set of random variables $\{X_{1},\ldots,X_{n}\}$ whose causal relationships are described by a directed acyclic graph (DAG) $\mathcal{G}$, a SCM with respect to the DAG $\mathcal{G}$ is a set of equations of the form $X_{i}=f_{i}(\boldsymbol{X}_{pa{(i)}},N_{i}), i=1,\ldots,n$, where $pa(i)$ denotes the set of direct parents of the node $i$ in $\mathcal{G}$, and the noise variables $N_{1},\ldots,N_{n}$ are jointly independent. 

% the function form  
% \begin{equation*}
% 	X_{t}^{i} = f_{i}(X^{i}_{t},\ldots,X^{i}_{t-p},\boldsymbol{X}^{pa(i)}_{t},\ldots,\boldsymbol{X}^{pa(i)}_{t-p},N^{i}_{t}),
% \end{equation*} 
% where the focus is on

In light of the ubiquity of time series data, it is appealing to adapt the results for i.i.d. data to \emph{stationary} time-dependent data. The ANMs have been extended to \emph{stationary} time series data. In~\cite{peters2013causal}, the \emph{time series models with independent noise} (TiMINo) considers 
 time-invariant functional relationships and \iid noise. Even though the processes generated according to TiMINo are not necessarily stationary, the stationarity of the data is required for the estimation  procedure. The well-known Granger causality is designed for vector autoregressive (VAR) models~\cite{granger1969investigating} without considering instantaneous effects, while LiNGAM-t~\cite{hyvarinen2010estimation} incorporates instantaneous effects and non-Gaussian noise. A bivariate \emph{deterministic} model via a \emph{linear time-invariant} filter is studied in~\cite{shajarisales2015telling}. The directed information rate~\cite{massey1990causality,kramer1998directed} from information theory is defined for bivariate stationary processes (see~\cite{amblard2013relation} for its relationship with Granger causality). 
 
%, where the identification proof of TCM is built on LiNGAM and ANMs, and CD-NOD is a constraint-based method~\cite{spirtes2000causation}. 
 
 There are a few works on causal inference through the lens of nonstationarity~\cite{huang2015identification,Huang0GG19,huang2020causal}. The \emph{time-dependent causal model}~\cite{huang2015identification} (referred to as TCM in this paper) and CD-NOD~\cite{huang2020causal} model nonstationarity by introducing a surrogate random variable to represent time. TCM deals with time-dependent functional relationships, but the estimation procedure becomes more challenging due to the nonstationarity of the data. The authors in~\cite{Huang0GG19} study a linear model (where the coefficients follow the autoregressive models) with additive noise that are uncorrelated in time, and the estimation step relies on nonlinear state-space model estimation procedures. However, none of them is built on well-established frameworks for \emph{slowly varying nonstationary processes} such as evolutionary spectra~\cite{Priestley1965}, Wigner-Ville spectral analysis~\cite{martin1985}, and locally stationary processes~\cite{Dahlhaus1996a} among others, from which powerful estimation procedures could be borrowed to greatly facilitate causal discovery tasks. In this work, we attempt to bridge this gap by proposing a new class of restricted SCMs that allows causal structure identification in generic cases and can be reliably estimated leveraging the bivariate evolutionary spectra framework~\cite{priestley1973,rao1972test}. 
 
 Our contribution is threefold. First,  we focus on a class of processes generated by \emph{linear time-varying filters} along with \emph{stationary Gaussian noise}, and develop theoretical results showing that the causal direction is identifiable in generic cases by exploiting the nonstationarity of the data. It is worth stressing that our framework can deal with instantaneous effects, which is an appealing property in comparison with Granger causality. Second, we extend these results to a network setting via a directed acyclic graph (DAG), where the processes are connected through time-varying linear relationships and the root nodes are assumed to be stationary. The identification result again relies on nonstationarity and this is in contrast to existing works where non-Gaussianity~\cite{shimizu2006linear} or nonlinearity~\cite{peters2014causal} is required for identification. Third, we develop efficient estimation algorithms, leveraging a recent variant of the evolutionary spectra estimate~\cite{xiang2019}, that perform well on a variety of synthetic and real datasets, including challenging ones with non-smooth and high-order filters.

The paper is organized as follows. In Section~\ref{sec:iden}, we present our main result on causal identification of a nonstationary bivariate linear model with time-varying coefficients. Various properties of the time-varying lag operator are discussed. In Section~\ref{sec:estimate}, we present our causal inference procedure, building on the bivariate evolutionary spectra estimates and stationarity test. We extend these results to a network setting in Section~IV and report our experimental results in Section~\ref{sec:exp}. 

%Preliminary results on the bivariate case have been reported in~\cite{Du2021}. In this extended version, (1) we study in greater depth the bivariate case through various concrete settings, and  more importantly, (2) we formulate a network setting and develop identification results as well as efficient estimation procedures, supported empirically by the superior performance of our methods in various simulation results. 

\subsection{Notation}

Let $\mathbb{Z}$, $\mathbb{Z}_{\geq 0}$, and $\mathbb{C}$ denote the integers, non-negative integers, and complex numbers, respectively. We use $\bar{\mathbb{Z}}_{\geq 0}$ to denote $\mathbb{Z}_{\geq 0} \cup \{\infty\}$. A sequence of random variables is denoted by $\{X_{t}\} \triangleq \{X_{t}, t \in \mathbb{Z}\}$ with mean function $\mu_{X,t}=\E[X_t]$ and auto-covariance function $\gamma_{XX}(r,s)=\Cov(X_r,X_s)$. We write $\{X_{t}\} \independent \{Y_{t}\}$ to denote the (statistical) independence between $\{X_{t}\}$ and $\{Y_{t}\}$, which requires the random vectors $(X_{t_{1}},\ldots,X_{t_{n}})$ and $(Y_{t_{1}},\ldots,Y_{t_{n}})$ to be independent for any $n>0$ and any sequence $t_{1},\ldots,t_{n} \in \mathbb{Z}$. Throughout this work, \emph{stationary} processes is referred to as \emph{wide-sense stationary} processes. We use the capital Greek letter (e.g., $\Phi, \Psi$, H) for polynomial function and the corresponding lower case (i.e., $\phi, \psi$, $\eta$) for its coefficients. For a matrix $A\in \mathbb{C}^{n \times n}$, we use $|A|$ and $\textstyle||A||_p \triangleq \max_{x \neq0} (||Ax||_{p}/||x||_{p})$ , $p\geq 1$, to denote its determinant and the matrix norm induced by the $l_{p}$ norm, respectively.  We use $\rho(A) \triangleq \max_{1\leq j \leq n}|\lambda_{j}|$ to denote the spectral radius of the matrix $A$, where $\{\lambda_{1},\ldots,\lambda_{n}\}$ are the eigenvalues of $A$. %Throughout this paper, we use $C$ to denote a genetic constant. 

%\section{Related Work}

 \section{Model identifiability in the bi-variate case}
 \label{sec:iden}
 \subsection{Linear time-varying filter with additive stationary noise}
\label{sec:lag}
For a process $\{X_{t}\}$, we define the \emph{lag operator} $\mathsf{B}$ as $\mathsf{B}^{j}X_{t} \triangleq X_{t-j}, j \in \bar{\mathbb{Z}}_{\geq 0}$. Let $\textstyle\Phi_{t}^{p}(z) \triangleq \sum_{j=0}^{p} \phi_{t,j}z^{j},z \in \mathbb{C},$ denote a time-dependent polynomial function of finite degree $p$. If $\Phi_{t}^{p}(z)$ is not constantly zero, we require $\phi_{t,p} \neq 0$ for some $t$. For infinite degree $p=\infty$, we define $\textstyle\Phi_{t}^{\infty}(z) \triangleq \sum_{j=0}^{\infty} \phi_{t,j}z^{j}$, with $z \in \mathbb{C}$ such that $|z| \leq 1$, where the coefficients of $\Phi_{t}^{\infty}(z)$ are assumed to be \emph{absolutely summable}, i.e., $ \textstyle\sum_{j=0}^{\infty} |\phi_{t,j}|<\infty$. Given a polynomial function  $\Phi_{t}^{p}(z)$ of degree $p \in \bar{\mathbb{Z}}_{\geq 0}$, a \emph{time-varying (lag-polynomial) operator} of order $p$ is defined as
\begin{equation}
    \Phi_{t}^{p}(\mathsf{B}) \triangleq \sum_{j=0}^{p}\phi_{t,j}\mathsf{B}^{j}.
    \label{eq_def1}
\end{equation}
We call an operator time-invariant if it does not dependent on $t$. For an operator $\Phi_{t}^{p}(\mathsf{B})$ of finite order $p$, if there exists an operator $\Theta_{t}^{r}(\mathsf{B}), r \in \bar{\mathbb{Z}}_{\geq 0}$, such that $\Theta_{t}^{r}(\mathsf{B})\Phi_{t}^{p}(\mathsf{B}) = 1$, we call $\Theta_{t}^{r}(\mathsf{B})$ the \emph{(left) inverse operator} of $\Phi_{t}^{p}(\mathsf{B})$, which is denoted by $(\Phi_{t}^{p}(\mathsf{B}))^{-1}$. 

In this work, we start with a class of bivariate Gaussian processes $\{X_{t},Y_{t}\}$ that are trend free ($\mu_{X,t}=\mu_{Y,t}=0$) and follow the following model
\begin{align}
    \label{eq_forw}
    Y_{t} =  \Phi_{t}^{p}(\mathsf{B})X_{t} + N_{t}, \quad p \in \mathbb{Z}_{\geq 0}, \quad \{N_{t}\} \independent \{X_{t}\},      
\end{align}
where the noise $\{N_{t}\}$ is a stationary process, and we assume that $\Phi_{t}^{p}(\mathsf{B})$ is invertible.   The assumption that  $\Phi_{t}^{p}(\mathsf{B})$ is invertible (see Lemma~\ref{lem:inverse} for details) implies that our model always includes the instantaneous effects (i.e., $\phi_{t,0}\neq 0$), which is regarded as a more difficult case compared with the one without instantaneous effects~\cite{hyvarinen2010estimation}. Also note that~\eqref{eq_forw} can be equivalently written as any invertible time-invariant operator applied to its both sides (since the noise remains stationary). We will thus focus on the representation in~\eqref{eq_forw} for simplicity.
\begin{remark}	 
	 We do not specify the generating process of $\{X_{t}\}$, which is in contrast to the bivariate version of SCMs in~\cite{hyvarinen2010estimation, peters2013causal, huang2015identification,Huang0GG19} where the cause is assumed to be a noise variable. Our bivariate setting is more challenging in that one could use a stationarity test to tell apart the cause from effect if $\{X_{t}\}$ is always stationary. In our network setting in Section~\ref{sec:network}, however, we will have to assume the root nodes are stationary, since the problem seems to be intractable otherwise.  
	 \end{remark}

\smallskip
 We say a backward model exists if there exists $\Psi_{t}^{q}(\mathsf{B}), q \in \bar{\mathbb{Z}}_{\geq 0}$, and a stationary process $\{\widetilde{N}_{t}\}$ such that 
\begin{equation}
    \label{eq_back}
    X_{t} =  \Psi_{t}^{q}(\mathsf{B})Y_{t} + \widetilde{N}_{t}, \quad \{\widetilde{N}_{t}\} \independent \{Y_{t}\}.
\end{equation}
The \emph{causal direction} $x \to y$ is said to be \emph{identifiable} if the joint distribution of $\{X_{t},Y_{t}\}$ does not admit a backward model~\eqref{eq_back}. Note that  a valid backward model requires the coefficients of $\Psi_{t}^{q}(\mathsf{B})$, i.e., $\{\psi_{t,i}\}$ to be absolutely summable. 
%
%In the next section, we will first present the  identifiability results of model~\eqref{eq_forw} for the general setting in Theorem~\ref{thm:identification} .
%

% Even though the model in~\eqref{eq_forw} does not allow lag terms for $\{Y_t\}$, its analysis is highly non-trivial, as we shall see in this section. We thus leave the extension of this model for future work. 

\subsection{Identifiability}
\label{sec:identifia}

Recall that for bivariate Gaussian processes, the backward model defined in~\eqref{eq_back} has to satisfy two constraints: \emph{the independence constraint} ($\{\widetilde{N}_{t}\} \independent \{Y_{t}\}$) 
	and \emph{the stationarity constraint} ($\{\widetilde{N}_{t}\}$ is stationary). Our main theorem characterizes two necessary conditions, corresponding to these two constraints, regarding the existence of a backward model~\eqref{eq_back}. To illustrate that the constraints for a backward model to exist are hard to be satisfied, we provide the identifiability results for the \iid setting in Corollary~\ref{coro:iidsetting} and~\ref{coro:zero_order}. 
%\smallskip

\begin{theorem}
\label{thm:identification}
Let $\{X_{t}, Y_{t}\}$ be a bivariate Gaussian process following the model~\eqref{eq_forw} such that
\begin{equation}
\label{defYt}
     Y_{t} = \Phi_{t}^{p}(\mathsf{B})X_{t} + N_{t},\end{equation}
where we assume that $\Phi_{t}^{p}(\mathsf{B})$ is invertible. Then a backward model of \eqref{defYt} exists only if the following two conditions are satisfied. 
\begin{enumerate}[(I)]
	\item\emph{Condition for the independence constraint.}  The equation $\gamma_{XX}(t_{1},t_{2}) = H_{t_{2}}^{s}(\mathsf{B}) \alpha(t_{1},t_{2})$ with respect to $H_{t_{2}}^{s}(\mathsf{B})$, where
\begin{align*}
        &\alpha(t_{1},t_{2}) = \Phi_{t_{2}}^{p}(\mathsf{B})\gamma_{XX}(t_{1},t_{2}) \\
        &\hspace{7em}+ (\Phi_{t_{1}}^{p}(\mathsf{B}))^{-1}\gamma_{NN}(t_{2}-t_{1}), 
\end{align*}
determines a nonempty class of operators $\mathcal{O}$ such that for any $H_{t}^{s}(\mathsf{B}) \in \mathcal{O}$, $\{Y_{t}\} \independent \{X_{t}-H_{t}^{s}(\mathsf{B})Y_{t}\}$, and $\{H_{t}^{s}(\mathsf{B})Y_{t}\}$ has a unique distribution. 
\item \emph{Condition for the stationarity constraint.} Let $\Theta_{t}^{r}(\mathsf{B})\triangleq (\Phi_{t}^{p}(\mathsf{B}))^{-1}$, then there exists an operator $H_{t}^{s}(\mathsf{B}) = \sum_{j=0}^{s}\eta_{t,j}\mathsf{B}^{j}$ in $\mathcal{O}$ such that
 \begin{equation*}
     \sum_{j=1}^{s}\sum_{k=1}^{r}\eta_{t,j}\theta_{t-j,k}\gamma_{NN}(k-j)
 \end{equation*} 
is \emph{time-invariant}.
\end{enumerate}
 
\end{theorem}
\smallskip

The proof of Theorem~\ref{thm:identification} is provided in Appendix~\ref{sec:thm1}. As shown in the proof of Theorem~\ref{thm:identification}, the second condition is a consequence of the first one. However, the second condition itself is quite strong in that a combination of time-varying coefficients has to be time-invariant. As a result, Theorem~\ref{thm:identification} implies that the causal direction is likely to be identifiable in generic cases, which is further supported by our experimental results on both synthetic and real-world datasets in Section~\ref{sec:exp}. Note that this is analogous to the identifiability results for the nonlinear ANMs~\cite{hoyer2009nonlinear} where the backward model only exists under strong conditions. Estimating the time-varying coefficients remains a challenging task. Fortunately, reliable estimation procedures are available for a class of slowly varying processes, called bivariate evolutionary spectra processes~\cite{priestley1973,rao1972test}, based on which we propose a natural causal discovery procedure in Section~\ref{sec:estimate}.

As a consequence of the two constraints, we have the following corollary when both $\{X_{t}\}$ and $\{N_{t}\}$ are \iid Gaussian processes (see Appendix~\ref{sec:coro1} for the proof).

\smallskip
\begin{corollary}
	\label{coro:iidsetting}
If $\{X_{t}\}$ and $\{N_{t}\}$ are two \iid Gaussian processes with $\E[X_t^2]=\s_X^2$ and $\E[N_t^2]=\s_N^2$, and $\phi_{t,p}\neq 0$ for all $t$, then the coefficients of the operator $\Psi^{q}_{t}(\mathsf{B})$ in \eqref{eq_back} are determined by $\psi_{t,0} = 1/\phi_{t,0}$ and  
%\begin{align}
%\psi_{t,i} &= \frac{\phi_{t+p-i,p-i}}{\phi_{t+p-i,p}\phi_{t-i,0}} -\nonumber\\
%&\frac{1}{\sigma_{X}^{2}\phi_{t+p-i,p}\phi_{t-i,0}} \sum_{j=\max(0,i-2p)}^{i-1}\psi_{t,j}\E[Y_{t-j}Y_{t+p-i}],
%\label{lemma2_solve}
%\end{align}
\begin{align}
\psi_{t,i} =  \frac{-1}{\phi_{t-i,0}}\left(\sum_{j = 1}^{\min(p,i)}\psi_{t,i-j}\phi_{t-i+j,j} + \frac{\sigma_{N}^{2}\psi_{t,i-p}}{\sigma_{X}^{2}\phi_{t+p-i,p}}\right)
\label{lemma2_solve}
\end{align}
for $i \geq 1$. A backward model~\eqref{eq_back} exists only if  $\{\psi_{t,i}\}$ is absolutely summable and $
\Var(\widetilde{N}_{t})=\sum_{j=1}^{\min(s,r)}\psi_{t,j}\theta_{t-j,j}\gamma_{NN}(0)$ is time-invariant, where $\{\theta_{t,j}\}$ are the coefficients of the inverse operator of $\Phi^{p}_{t}(\mathsf{B})$ (see equation~\eqref{sol_coeff} below).
\end{corollary}
\smallskip

\begin{remark}
As the SNR $ \sigma_{X}^{2}/ \sigma_{N}^{2}$ goes to infinity, the coefficients $\{\psi_{t,i}\}$, converges to $\theta_{t,0} = 1/ \phi_{t,0}$ and
\begin{equation}
   \theta_{t,i} =  -\frac{1}{\phi_{t-i,0}}\sum_{j = 1}^{\min(p,i)}\theta_{t,i-j}\phi_{t-i+j,j}, \quad i \geq 1,
   \label{sol_coeff}
\end{equation}
which are the coefficients of $\Theta_{t}^{r}(\mathsf{B}) \triangleq (\Phi^{p}_{t}(\mathsf{B}))^{-1}$ (see the derivation of~\eqref{sol_coeff} in~\cite[equation~(4.10)]{abdrabbo1967prediction}). Thus the invertibility of $\Phi_{t}^{p}(\mathsf{B})$ is a necessary condition for a backward model to exist when the SNR is sufficiently high. Since we assume that $\Phi^{p}_{t}(\mathsf{B})$ is invertible, we thus focus on the cases when the identifiability is more difficult to show.
\end{remark}
\smallskip

In Corollary~\ref{coro:iidsetting}, we show that $\{\psi_{t,i}\}$ can be solved iteratively, and the variance of $\{\widetilde{N}_{t}\}$ is written as a combination of $\{\psi_{t,i}\}$ and $\{\theta_{t,i}\}$. In general, it could be hard to check whether $\{\psi_{t,i}\}$ is absolutely summable and whether $\{\widetilde{N}_{t}\}$ is stationary. To get a concrete sense of the identifiability result, we simplify the setting by letting $\Phi_{t}^{p}(\mathsf{B})$ to be of zero order in the following corollary (see Appendix~\ref{sec:coro2} for the proof). 
\smallskip

\begin{corollary}
	\label{coro:zero_order}
	Let $\{X_{t}\}$ and $\{N_{t}\}$ be \iid Gaussian processes with zero means and variances $\s_X^{2}$ and $\s_N^{2}$, respectively. Consider the following forward model with $\phi(t) \neq 0$,
	\begin{equation}
	\label{model_prop2}
		Y_{t} = \phi(t)X_{t} + N_{t}, \quad \{ {N}_{t}\} \independent \{X_{t}\}.
	\end{equation}
Then there exists a model as follows,
	\begin{equation}
	\label{back_prop1}
		X_{t} = \frac{\phi(t)}{\phi^{2}(t)+\sigma_{N}^{2}/\sigma_{X}^{2}}Y_{t} + \widetilde{N}_{t},  \quad \{ \widetilde{N}_{t}\} \independent \{Y_{t}\},
	\end{equation}
	where $\{\widetilde{N}_{t}\}$ is determined by $\widetilde{N}_{t}= \frac{\sigma_{N}}{\sqrt{\phi^{2}(t)+\sigma_{N}^{2}/\sigma_{X}^{2}}}  W_{t}$, where $\{W_{t}\}$ is an \iid process with $\sigma_{W}^{2}=1$.
\end{corollary}
	
	\smallskip
\begin{remark}
	Due to the stationarity constraint on $\{\widetilde{N}_{t}\}$, a backward model exists only if $|\phi(t)|$ is time-invariant. The noise $\{\widetilde{N}_{t}\}$ has the form of a stationary process multiplied by a nonnegative function, which belongs to a class of nonstationary processes call the uniformly modulated process (UMP)~\cite{Priestley1965} (see the definition of UMP in Section~\ref{sec:estimate}).
	 \end{remark}
	 
\smallskip	 

If the stationary noise assumption is relaxed to be the UMP noise, then a backward model always exists in the setting of Corollary~\ref{coro:zero_order}. But in the general setting, by replacing $Y_{t}$ in~\eqref{eq_back} with~\eqref{eq_forw}, one can write \begin{equation}
	\widetilde{N}_{t} = (1- \Psi_{t}^{q}(\mathsf{B})\Phi_{t}^{p}(\mathsf{B}) ) X_{t} - \Psi_{t}^{q}(\mathsf{B})N_{t}, \nonumber
\end{equation}
which is a sum of two independent processes. In generic cases, $\{\widetilde{N}_{t}\}$ is not only nonstationary but non-UMP. Thus our model is likely to be identifiable even if we consider the UMP noise. This is also supported empirically by our experimental results on synthetic data in Section~\ref{sec:exp}.

\subsection{Time-varying operator}
\label{sec:operator}
In order to establish the identifiability results of our model, we need to first investigate some key properties of the time varying operator. 
We say an operator $\Phi_{t}^{p}(\mathsf{B})$ is \emph{time-invariant} if $\phi_{t,j} = \phi_{t-1,j}$ holds for all $j \geq 0$ and $t \in \mathbb{Z}$. By applying the operator $\Phi_{t}^{p}(\mathsf{B})$ to $\{X_{t}\}$, we obtain
\begin{equation}
\label{eq_apply}
   \Phi^{p}_{t}(\mathsf{B}) X_{t} =  \sum_{j=0}^{p}\phi_{t,j}X_{t-j}.
\end{equation}

Since we focus on Gaussian processes and operators with absolutely summable coefficients, we would like to have any series of the form in~\eqref{eq_apply} to converge even when $p=\infty$. To address this technical issue, we present the following proposition, and the proof of which is a straightforward extension of the time-invariant case proved in~\cite{davis1987time} (and we include it in Appendix~\ref{app:prop1} for completeness).
 
\smallskip
\begin{proposition}
\label{prop:infty}
Let $\{X_{t}\}$ be a sequence of random variables such that $\sup_{t}\E[|X_{t}|]<\infty$. If $\textstyle\sum_{j=0}^{\infty}|\psi_{t,j}|<\infty$, then the series
\begin{equation}
    \label{eq_prop1}
    \Psi_{t}^{\infty}(\mathsf{B})X_{t} =  \sum_{j=0}^{\infty}\psi_{t,j}X_{t-j}
\end{equation}
converges absolutely with probability one. If $\sup_{t}\E[|X_{t}|^{2}] < \infty$, the series converges in mean square to the same limit.
\end{proposition} 
\smallskip

 We now discuss the relationship between different operators. First, we say two operators $\textstyle\Phi_{t}^{p}(\mathsf{B})$ and $\textstyle\Psi_{t}^{q}(\mathsf{B})$, with $p, q \in \bar{\mathbb{Z}}_{\geq 0}$, are equivalent if $\phi_{t,j}=\psi_{t,j}$ holds for all $j \geq 0$ and $t \in \mathbb{Z}$, and we write $\Phi_{t}^{p}(\mathsf{B}) = \Psi_{t}^{q}(\mathsf{B})$. Otherwise, we use $\neq$ to denote they are not equivalent. To facilitate the analysis, we will make use of an equivalent definition for the rest of the paper. We write $\Phi_{t}^{p}(\mathsf{B}) = \Psi_{t}^{q}(\mathsf{B})$ if 
 	\begin{equation}
 	 	\label{eq_poly_equal}
   \Phi_{t}^{p}(z) =  \Psi_{t}^{q}(z)
\end{equation}
holds for $z$ in some open set $\mathbb{E} \subseteq{C}$ that contains $0$ (see Appendix~\ref{app:def} for the proof of equivalence).

It can be easily verified that time-varying lag-polynomial operators do not satisfy the commutative property of multiplication in general, i.e.,
\begin{equation}
    \Phi_{t}^{p}(\mathsf{B})\Psi_{t}^{q}(\mathsf{B}) \not = \Psi_{t}^{q}(\mathsf{B})\Phi_{t}^{p}(\mathsf{B}).\label{eq:noncommutative}
\end{equation}

\begin{remark}
	For two operators $\Phi_{t_{1}}^{p}(\mathsf{B})$ and $\Psi_{t_{2}}^{q}(\mathsf{B})$ such that $t_{1}$ and $t_{2}$ do not depend on each other, we have $\Phi_{t_{1}}^{p}(\mathsf{B})\Psi_{t_{2}}^{q}(\mathsf{B}) = \Psi_{t_{1}}^{q}(\mathsf{B})\Phi_{t_{2}}^{p}(\mathsf{B})$.
\end{remark} 

\smallskip

% Due to the definition of the equivalence of two operators, the existence of the inverse operator $\Theta_{t}^{r}(\mathsf{B})$ is equivalent to the following fact:  The equation $\sum_{k=0}^{r}\sum_{j=0}^{p}\theta_{t,k}\phi_{t-k,j}z^{j+k} = 1$, $z$ in some open set $\mathbb{E} \subseteq{C}$ that contains $0$, and has a solution $\{\theta_{t,j}\}$ that is absolutely summable. 
As a consequence of the non-commutative property in~\eqref{eq:noncommutative}, $\Phi_{t}^{p}(\mathsf{B})$ may not be the inverse operator of $ (\Phi_{t}^{p}(\mathsf{B}))^{-1}$ in general. It is known that, when $\Phi_{t}^{p}(\mathsf{B})$ is time-invariant, a necessary and sufficient condition for $\Phi^{p}(\mathsf{B})$ to be invertible is
\begin{equation}
    \label{time_inva_inverse}
    \sum_{j=0}^{p}\phi_{j}z^{j} \neq 0, 
\end{equation}
for $|z| \leq 1$, which says that the roots of the polynomial in \eqref{time_inva_inverse} are strictly outside the unit circle. A similar statement was proved in~\cite{davis1987time}.  When $\Phi_{t}^{p}(\mathsf{B})$ is time-varying and $\phi_{t,p}\neq 0$ for all $t$, a necessary and sufficient condition for the invertibility of $\Phi^{p}_{t}(\mathsf{B})$ is provided in~\cite{hallin1986non} using Green's functions, making the evaluation of the condition very challenging.   In the following, we provide two sufficient conditions and one necessary condition on the existence of inverse operators that are easy to check. Some examples will be discussed afterwards to illustrate the conditions.

 %In equation~\eqref{sol_coeff}, we see that the coefficients $\{\theta_{t,i}\}$ can be solved iteratively. Then, in order for the inverse operator $\Theta_{t}^{r}(\mathsf{B})$ to be well defined, the coefficients $\{\theta_{t,i}\}$ has to be absolutely summable.
 
 %We start with a simple proposition to show that the existence of an inverse operator implies its uniqueness (see Appendix~\ref{app:unique} for the proof). 

\begin{comment} 
\smallskip
\begin{proposition}
\label{prop:unique}
Let $\Phi_{t}^{p}(\mathsf{B})$ be an operator of finite order $p$. If there exists $\Theta_{t}^{r}(\mathsf{B}), r \in \bar{\mathbb{Z}}_{\geq 0}$, that is an inverse operator of $\Phi_{t}^{p}(\mathsf{B})$, then $\Theta_{t}^{r}(\mathsf{B})$ is unique.
\end{proposition}
\smallskip
\end{comment}

\smallskip
\begin{lemma}
\label{lem:inverse}
Let $\Phi_{t}^{p}(\mathsf{B})$ be an operator of finite order $p \geq 1$, and we assume that $\phi_{t,j}\neq 0$ for some $j \geq 0$ for each $t$. 
\begin{enumerate}[(I)]
\item \emph{Sufficient conditions for $\Phi_{t}^{p}(\mathsf{B})$ to be invertible}.\\ 
The inverse operator $(\Phi_{t}^{p}(\mathsf{B}))^{-1}$
exists if either of the following conditions holds,
%\begin{enumerate}[(a)]
%\item $\label{lemma_condi_a} |\phi_{t,0}|> \sum_{j=1}^{p}|\phi_{t+j,j}|>0$.
%\item $\label{lemma_condi_b}
%      \phi_{t,0} > \phi_{t+1,1} > \ldots > \phi_{t+p,p} \geq 0$.
%      
\begin{align}
&\text{(a) }\quad |\phi_{t,0}|> \sum_{j=1}^{p}|\phi_{t+j,j}|>0,\label{lemma_condi_a}\\
&\text{(b) }\quad\phi_{t,0} > \phi_{t+1,1} > \ldots > \phi_{t+p,p} \geq 0.\label{lemma_condi_b}
\end{align}
%\end{enumerate}

\item \emph{Necessary condition for $(\Phi_{t}^{p}(\mathsf{B}))^{-1}$ to have a finite order}. 

$\Phi_{t}^{p}(\mathsf{B})$ has an inverse operator of finite order $q$ only if $\phi_{t,0} \neq 0$ and 
\begin{equation}
    \prod_{i=0}^{q}\phi_{t-i,p}=0.
    \label{nece}
\end{equation}

\end{enumerate}
\end{lemma}
\smallskip
 
\begin{remark}
If $\Phi_{t}^{p}(\mathsf{B})$ is time-invariant, then \eqref{nece} reduces to $\phi_{t,p}=0$ for all $t$, which contradicts the definition of $\Phi_{t}^{p}(\mathsf{B})$ since it requires that $\phi_{t,p}\ne 0$ for some $t$. Thus the inverse operator of  $\Phi_{t}^{p}(\mathsf{B})$ cannot be of finite order in the time-invariant case.
\end{remark}
\smallskip

%\begin{remark}
%If $\Phi_{t}^{p}(\mathsf{B})\label{rmk:2}$ is of order 1, then $|\phi_{t+1,1}/\phi_{t,0}|<1$, with $\phi_{t,0}\neq 0$, is a sufficient condition for the existence of an inverse operator. Since \eqref{sol_coeff} becomes $\textstyle \psi_{t,i} =  -(\phi_{t-i+1,1}/ \phi_{t-i,0})\psi_{t,i-1}$, then there exists $0<\varepsilon<1$ such that $|\psi_{t,i}| \leq |\phi_{t+1,1}/\phi_{t,0}| |\psi_{t,i-1}| \leq\varepsilon|\psi_{t,i-1}|$, it follows that $\sum_{j=0}^{\infty}|\psi_{t,i}|\leq |\psi_{t,0}|\sum_{j=0}^{\infty}\varepsilon^{j}<\infty$.
%\end{remark}
%\smallskip
The proof of Lemma~\ref{lem:inverse} is provided in Appendix~\ref{app:lemma1}. In Section~\ref{est_transf}, we show that there is a close relationship between condition~\eqref{lemma_condi_a} and a slowly varying condition on the coefficients from the evolutionary spectra framework. The necessary condition in Lemma~\ref{lem:inverse} says that an inverse operator of finite order exists only if $\phi_{t,p}=0$ for infinitely many $t$. This condition characterizes a class of operators that could be restrictive since it does not contain the time-invariant operators (and recall that the inverse operator of a time-invariant operator is of infinite order). We therefore consider the ``complement'' of this class to be a more general class of operators. 

% Since conditions \eqref{lemma_condi_a} and \eqref{lemma_condi_b} are each a sufficient condition for $\Phi^{p}(\mathsf{B})$ to be invertible, we obtain the following two sufficient conditions for \eqref{time_inva_inverse} to hold $\text{(a) }|\phi_{0}|> \sum_{j=1}^{p}|\phi_{j}|>0$ or $\text{(b) }\phi_{0} > \phi_{1} > \ldots > \phi_{p} > 0$.

%In both finite- and infinite-order cases, the coefficients of the inverse operator $\Psi_{t}^{q}(\mathsf{B}) \triangleq (\Phi_{t}^{p}(\mathsf{B}))^{-1} , q \in \bar{\mathbb{Z}}_{\geq 0}$, are determined by $\textstyle \psi_{t,0}\phi_{t,0}=1$ and 
%\begin{equation}
%   \psi_{t,i} =  -\frac{1}{\phi_{t-i,0}}\sum_{j = 1}^{\min(p,i)}\psi_{t,i-j}\phi_{t-i+j,j}, \quad i \geq 1.
%\end{equation}
%See details of this claim in Appendix~\ref{app:lemma1}.

 Now we provide three examples to show that inverse operators exist under the conditions in Lemma~\ref{lem:inverse}.

\smallskip
{\noindent \bf Example 1.}
Consider the first-order operator $\Phi^{1}_{t}(\mathsf{B}) \triangleq 1+\phi_{t,1}\mathsf{B}$, where $\phi_{t,1} = 1$ when $t$ is even, and $\phi_{t,1} = 0$ otherwise. One can check that condition \eqref{nece} holds for all $q \geq 1$. Then using \eqref{sol_coeff}, it is straightforward to find that 
\begin{equation*}
   (1-\phi_{t,1}\mathsf{B})(1+\phi_{t,1}\mathsf{B}) = 1 - \phi_{t,1}\phi_{t-1,1}\mathsf{B}^{2} = 1.
\end{equation*}

{\noindent \bf Example 2.}
Consider the first-order operator $\Psi^{1}_{t}(\mathsf{B}) \triangleq 1+\psi_{t,1}\mathsf{B}$, with $0<|\psi_{t,1}|<1$. One can check that condition \eqref{nece} does not hold for any $q\geq 1$ while condition \eqref{lemma_condi_a} holds immediately. Then by \eqref{sol_coeff}, we obtain
\begin{equation}
   (1 +\psi_{t,1}\mathsf{B})^{-1} = 1
   +\sum_{j=1}^{\infty}\biggl((-1)^{j}\prod_{k=1}^{j}\psi_{t-k+1,1}\biggr) \mathsf{B}^{j}.
   \label{1st_inverse}
\end{equation}
\smallskip
{\noindent \bf Example 3.}
Consider the operator $\Psi^{1}_{t}(\mathsf{B})$ in {\bf Example 2.} with $\psi_{t,0}=1$ and $\psi_{t,1} = 0.5\cos(t/T), T \in \mathbb{Z}$. Since $t/T$ is a rational number for any $t \in \mathbb{Z}$, we have $\psi_{t,1} \neq 0$ and $|\psi_{t,1}|<1$, which implies that \eqref{nece} does not hold for any $q \geq 1$. It then follows from~\eqref{lemma_condi_a} in Lemma~\ref{lem:inverse} that $\Psi^{1}_{t}(\mathsf{B})$ has an inverse operator of the form \eqref{1st_inverse}. This operator was employed in~\cite{priestley1973} as the transfer function for an open loop system.

\section{Model estimation}
\label{sec:estimate}
\subsection{Causal inference procedure}

To simplify the presentation of the estimation procedure in this section, we will adopt an alternative expression of the model~\eqref{eq_forw} without using the lag-polynomial operator. Consider a bivariate process $\{X_{t},Y_{t}\}$, we say the causal direction between $\{X_{t}\}$ and $\{Y_{t}\}$ is $x \to y$ if the following model holds,
\begin{equation}
    Y_{t} = \sum_{u=0}^{\infty} d_{t}(u) X_{t-u}+N_{t},\quad \{X_{t}\} \independent \{N_{t}\},
    \label{model_transfer}
\end{equation}
where $\{N_{t}\}$ is a stationary process, and $\{d_{t}(u)\}$ is called the time-varying filter. Conversely, if $\{X_{t},Y_{t}\}$ admits the model, $\textstyle  X_{t} = \sum_{u=0}^{\infty} \tilde{d}_{t}(u) Y_{t-u}+\widetilde{N}_{t}, \{Y_{t}\} \independent \{\widetilde{N}_{t}\}$, where $ \{\widetilde{N}_{t}\}$ is a stationary process, then we say the causal direction is $y \to x$. The assumptions on model~\eqref{model_transfer} that allow efficient estimation of $\{d_{t}(u)\}$ are technical and will be deferred to Section \ref{est_transf}, after a brief overview of the evolutionary spectra framework. We now describe our causal inference procedure in Algorithm~\ref{alg:1} to test the null hypothesis $H_0: x\to y$, and the test for $y\to x$ can be done in the same manner. Let $p_{I}^{x \to y}$ denotes the p-value from the independence test and $q_{S}^{x \to y}=1$ if the residual is stationary and $0$ if nonstationary. Similarly, we obtain $p_{I}^{y \to x}$ and $q_{S}^{y \to x}$ from the test for $y\to x$. We accept or reject $H_{0}$ by checking the following conditions. For a prefixed $\alpha$, we accept $H_0$ if $p_{I}^{x \to y} \geq \alpha$ and $p_{I}^{y \to x} <\alpha$. (Similarly, we reject $H_0$ if $p_{I}^{x \to y} < \alpha$ and $p_{I}^{y \to x} \ge \alpha$.) If $p_{I}^{x \to y} \geq \alpha$ and $p_{I}^{y \to x} \geq \alpha$, then we rely on the stationarity test: We accept $H_0$ if $q_{S}^{x \to y}=1$ and $q_{S}^{y \to x}=0$ (or reject $H_0$ if $q_{S}^{x \to y}=0$ and $q_{S}^{y \to x}=1$). The causal inference procedure remains undecided for all the other cases.    
%Similar to the ANMs framework~\cite{peters2014causal}, our algorithm cannot tell the causal direction if the models of the form \eqref{model_transfer} hold for both $x \to y$ and $y \to x$, or neither of the directions, for which we call the causal direction undecided. This may due to the violation of the model assumptions or the weakness of the proposed procedure.

%Briefly, there are two cases when the causal direction can be decided: 1. The independence test is not significant for only one direction; 2. The independence test and stationarity test are both not significant for only one direction. We adopt a kernel-based independence test developed for random processes~\cite{chwialkowski2014kernel} (and we will refer to it as HSICp) to test the independence between the hypothesized cause and the residuals.
Here are some comments regarding the implementation details in  Algorithm~\ref{alg:1}. Given a window size $N_{F}$, the maximal order of model \eqref{model_transfer} considered by our estimation procedure is $\lfloor N_{F}/2 \rfloor$ (see Section~\ref{sec:exp} for details). The order $p$ can be selected using AIC \cite{aic1973} or BIC \cite{schwarz1978estimating}. For a similar independence test task, previous works~\cite{peters2013causal,huang2015identification} have used a kernel-based independence test developed for \iid data~\cite{gretton2008kernel} (referred to as HSIC), which may suffer from high false positive rates in certain cases~\cite{chwialkowski2014kernel}. The estimation of the filter and the stationarity test are based on the evolutionary spectra framework~\cite{percival2020spectral} by incorporating the multitaper method as in the univariate case~\cite{constantine2011,xiang2019} (see more details below).

\subsection{Univariate nonstationary processes}

%In the evolutionary spectra framework~\cite{Priestley1965}, the main focus is the continuous-time setting, while the discrete-time setting follows immediately. 
%For simplicity of presentation, we focus on the discrete-time setting through this work. 
To set the stage, we start with a brief review of the evolutionary spectra framework~\cite{Priestley1965}. Consider a class of nonstationary processes $\{X_{t}\}$, with $\E[X_{t}]=0$ and $\E[|X_{t}|^2]<\infty$ for $t\in \Integer$, such that 
\begin{equation}
X_{t}=\int_{-\pi}^{\pi} \phi_t(w) dZ(w), t\in\Integer, \label{eq:evolutionary}
\end{equation}
for some family $\Fc$ of functions $\{\phi_t(w)\}$ (defined on $[-\pi, \pi]$ indexed by $t$) and a measure $\mu(w)$, where $Z(w)$ is an orthogonal increment process with $\E[|dZ(w)|^{2}]=d\mu(w)$. If there exists a family of functions $\Fc=\{\phi_t(w)=e^{iwt}A_t(w)\}$ such that $\{X_{t}\}$ can be represented as in~\eqref{eq:evolutionary}
%\begin{equation}
%X(t)=\int_{-\pi}^{\pi} e^{iw t}A_t(w) dZ(w), t\in\Integer, \label{oscillatory}
%\end{equation}
%
%\[
%\phi_t(w)=e^{iw t}A_t(w),
%\]
and for any fixed $w$, the Fourier transform of $h_w(t)\triangleq A_t(w)$ (viewed as a function of $t$), denoted by $H_w(v)$, 
%\begin{align*}
%h_w(t):=A_t(w)=\int_{-\pi}^{\pi} e^{it\t} H_w(v)dv
%\end{align*}
has an absolute maximum at the origin, then $\{X_{t}\}$ is called an {\em oscillatory process} with respect to {\em oscillatory functions} $\{e^{iwt}A_t(w)\}$, and the evolutionary spectrum at time $t$ with respect to $\Fc$ is 
\begin{align*}
	dF_t(w)=|A_t(w)|^2d\mu(w).
\end{align*}
Note that $h_w(t)= 1$ corresponds to the case when $\{X_{t}\}$ is a stationary process, which leads to $H_w(v)= \d(v)$, where $\d(\cdot)$ is the Dirac delta function. To estimate the evolutionary spectral density, Priestley~\cite{Priestley1965} proposed a double-window technique, consisting of a short-time Fourier transform and smoothing. Recently, the bias/variance/resolution tradeoff of a variant of the evolutionary spectra estimate, incorporating the multitaper method~\cite{thomson1982}, is characterized~\cite{xiang2019}. Interesting methodologies on neural processes can be found in~\cite{rupasinghe2020multitaper}.

%Throughout this paper, we assume that $\mu(w)$ is absolutely continuous with respect to Lebesgue measure. Thus the {\em evolutionary spectral density at time $t$} is 
%\[
%f_t(w)=|A_t(w)|^2\frac{d\mu(w)}{dw}.
%\]
%%Let $dH_w(\t)$ denote the Fourier transform of $A_t(w)$, i.e., 
%%\[
%%A_t(w)=\int_{-\pi}^{\pi} e^{it\t} dH_w(\t).
%%\]
% As mentioned above, for any fixed $w$, $H_w(v)$ is the Fourier transform of $h_w(t)$, i.e.,
%$H_w(v)=\sum_{t=-\infty}^{\infty} h_w(t)e^{-ivt}$. Let 
%\[
%B_{\Fc}(w)= \int_{-\pi}^{\pi} |v| |H_w(v)|dv,
%\]
%and each family $\Fc$ of oscillatory functions is called {\em semi-stationary} if $B_{\Fc}(w)$ is bounded for all $w$. Then $
%B_{\Fc}= \bigl(\sup_{w} B_{\Fc}(w)\bigr)^{-1}$ 
%is called the {\em characteristic width of $\Fc$}. A \emph{semi-stationary process} $\{X_{t}\}$ is defined as the one that can be represented as~\eqref{eq:evolutionary} with respect to a semi-stationary family $\Fc$. Let $\Cc$ denote the class of semi-stationary families such that $\{X_{t}\}$ can be represented as~\eqref{eq:evolutionary}. Then
%\begin{equation}
%	B_{X}= \sup_{\Fc\in \Cc} B_{\Fc}\label{B_X}
%\end{equation}
%is called the {\em characteristic width of $\{X_{t}\}$}. 
%

It is hard to characterize \emph{characteristic widths}~\cite{Priestley1965}, which quantifies the length of a ``stable" segment, exactly for semi-stationary processes~\cite{melard1989}. However, there is one important class of processes whose characteristic widths can be bounded from below. This class, termed as the {\em uniformly modulated processes (UMP)}~\cite{Priestley1965}, is of the following form: 
\begin{equation}
X_{t}=c(t)Y_{t}, \label{modulated}
\end{equation}
where $Y(t)$ is a stationary process with zero mean and spectral density $f_Y(w)$, and the Fourier transform of $c(t)$ has an absolute maximum at the origin. Thus it follows straightforwardly that 
\[
X_{t}=\int_{-\pi}^{\pi} c(t)e^{iw t} dZ(w),
\]
where $\E |dZ(w)|^2=dF_Y(w)$. The process introduced in \eqref{modulated} is an oscillatory process since $\Fc_Y=\{c(t)e^{iwt}\}$ is a family of oscillatory functions. The evolutionary spectrum with respect to $\Fc$ is $f_t(w)=c^2(t)f_Y(w)$. 

%The name, uniformly modulated processes, follows from the fact that for two different frequencies $w_1$ and $w_2$ in $[-\pi, \pi]$, the spectrum is modulated in the same way, i.e., $f_{t_1}(w_1)/f_{t_2}(w_1)=f_{t_1}(w_2)/f_{t_2}(w_2)$.

\begin{algorithm}[H]

    %\footnotesize
%        \captionsetup{font=footnotesize}
	    \caption{Causal inference procedure (bivariate)\label{alg:1}
}
	\begin{algorithmic}
    	\State {\textbf{Input:} $\{(X_{t},Y_{t})\}_{t=1}^{T}$}, window size $N_{F}$
    	\State {\textbf{Output:} 
    	\State \hskip2em p-values from the independence tests and decisions from the stationarity tests
    	}
\Procedure{test}{$x \to y$}
        \State  {\bf Estimation}:
         \State  \hskip1em   Estimate the filter $\hat{d}_{t}(u)$
        \State  \hskip1em Compute the residuals $\textstyle \hat{N}_{t} \triangleq Y_{t}- \sum_{u=0}^{p} \hat{d}_{t}(u) X_{t-u}$
        \State {\bf Independence test}:
         \State  \hskip1em Test whether $\{X_{t}\}\independent \{\hat{N}_{t}\}$        \State {\bf Stationarity test}:
            \State  \hskip1em Test the stationarity of  $\{\hat{N}_{t}\}$ \EndProcedure

\Procedure{test}{$y \to x$}
\EndProcedure
%\Procedure{test}{$y \to x$}\EndProcedure
%\If {one of $x \to y$ and $y \to x$ is possible}
%\State Conclude: $x \to y$ or $y \to x$
%\Else
% \State Conclude: undecided
%\EndIf
	\end{algorithmic}
	
\end{algorithm}

\subsection{Bivariate nonstationary processes}

Now we are ready to discuss bivariate processes $\{X_{t},Y_{t}\}$, consisting of two oscillatory processes, 
\begin{align*}
    X_{t}&=\int_{-\pi}^{\pi} A_{t,x}(w)e^{iwt} dZ_{x}(w),\\
    Y_{t}&=\int_{-\pi}^{\pi} A_{t,y}(w)e^{iwt} dZ_{y}(w),
\end{align*}
where $\{Z_{x}(w)\}$ with $\E|dZ_{x}(w)|^2=d\mu_{x}(w) $ and $\{Z_{y}(w)\}$ with $\E|dZ_{y}(w)|^2=d\mu_{y}(w)$ are two orthogonal increment processes, and  $\E[dZ_{x}(w)dZ_{y}^{*}(w)] \triangleq d\mu_{xy}(w)$. The evolutionary cross-spectrum~\cite{priestley1973} of $\{X_{t},Y_{t}\}$ at time $t$ with respect to $\mathcal{F}_{x}$ and $\mathcal{F}_{y}$ is
\begin{equation*}
    dF_{t,xy}(w) = A_{t,x}(w) A_{t,y}^{*}(w)d\mu_{xy}(w).
\end{equation*}
The cross-spectral density of $\{X_{t},Y_{t}\}$ at time $t$ is
\begin{equation*}
    f_{t,xy}(w) = A_{t,x}(w) A_{t,y}^{*}(w) \frac{d\mu_{xy}(w)}{dw}.
\end{equation*}
For $\{Y_{t}\} \triangleq \{X_{t}\}$, the cross-spectral density of $\{X_{t},Y_{t}\}$ reduces to the auto-spectral density of $\{X_{t}\}$. Note that $ f_{t,xy}(w)$ is in general a complex function. 
%One way of  estimating of the cross-spectral density $f_{t,xy}(w)$ is to make use of the estimation of the auto-spectral densities $f_{t,xx}(w)$ and $f_{t,yy}(w)$ ~\cite{xiang2019,priestley1973}. 
In this work, we adopt the multitaper method approach~\cite{xiang2019} for the estimation of the auto-spectral densities $f_{t,xx}(w)$ and $f_{t,yy}(w)$ and the cross-spectral density $f_{t,xy}(w)$.  

\subsection{Estimation of the filter}
\label{est_transf}
Following the model assumptions in~\cite{priestley1973}, we assume $\{X_{t}\}$ and $\{Y_{t}\}$ are semi-stationary processes. The filter $d_{t}(u)$ is assumed to satisfy the \emph{slowing-varying condition}~\cite{rao1972test} described as follows. Let $D_{t}(w)$ denote the Fourier transform of $d_{t}(u)$ with respect to $u$. For each $w$, consider $D_{t}(w)$ as a function of $t$, with (generalized) Fourier transform
\begin{equation*}
   D_{t}(w) = \sum_{\theta=-\infty}^{\infty} e^{i\theta t}  L_{w}(\theta). 
\end{equation*}
If $|L_{w}(\theta)|$ attains the maximum at $\theta=0$ for all $w$, i.e., $|L_{w}(0)|\geq |L_{w}(\theta)|$ for $\theta \neq 0$,  we say the slowly varying condition is satisfied. Since $t$ and $u$ in $d_{t}(u)$ 
are discrete, $L_w(\theta)$ is a periodic function of $w$ and $\theta$ both with period $2\pi$. It suffices to define the slowly varying condition in the period $-\pi< w,\theta \leq \pi$. By leveraging the powerful estimation procedure  in~\cite{priestley1973}, we estimate $d_{t}(u)$ by
\begin{equation*}
\hat{d}_{t}(u) =  \mathscr{F}^{-1}_{w}\{\hat{D}_{t}(w)\} =  \mathscr{F}^{-1}_{w} \biggl\{ \frac{\hat{f}_{t,yx}(w)}{\hat{f}_{t,xx}(w)}\biggr\},
\end{equation*}
where $ \mathscr{F}^{-1}_{w}$ denotes the inverse Fourier transform. 

 It is thus natural to ask whether the slow varying condition and either of the two sufficient conditions for invertibility in  Lemma~\ref{lem:inverse} can be satisfied simultaneously. We provide two classes of filters to show that this is indeed the case.
 
\smallskip
{\noindent \bf Example 4.}
Consider the filter $d_{t}(u)= \delta(u) + \sum_{k=1}^{l}1/2^{k}\cos(A_{k}t)\delta(u-k)$, where $A_{k} \in (-\pi,\pi), l \geq 1
$ and $\delta(u)$ is the unit impulse function\footnote{The unit impulse function is referred to as either the Dirac delta function $\delta(t)$ (with $\delta(0)=\infty$) for $t\in\Real$ or the unit sample function $\delta(n)$ (with $\delta(0)=1$) for $n\in\Integer$.}. Since we have $\sum_{k=1}^{l}1/2^{k}|\cos(A_{k}(t+k))|< \sum_{k=1}^{l}1/2^{k} <1$, the invertibility condition \eqref{lemma_condi_a} is satisfied. The functions $D_{t}(w)$ and $L_{w}(\theta)$ in  the period $-\pi< w, \theta  \leq \pi$ are given as follows,
\begin{align*}
	D_{t}(w) &= 1 + \sum_{k=1}^{l}\frac{1}{2^{k}}\cos(A_{k}t)e^{-ikw},\\
L_{w}(\theta) &= 2\pi\biggl(\delta(\theta)+\sum_{k=1}^{l}\frac{e^{-ikw}}{2^{k+1}}\left(\delta(\theta-A_{k})+\delta(\theta+A_{k})\right)\biggr). 
\end{align*}

Since $L_{w}(\theta)$ is a linear combination of delta functions, it is straightforward to see that the slowly varying condition is satisfied. 

\smallskip
{\noindent \bf Example 5.}  Consider the filter $d_{t}(u)$ with $d_{t}(0)=1$ and $\sum_{-\infty}^{\infty} |d_{t}(u)| <\infty$ for  $1 \leq  u \leq p$, for which we can assume that condition~\eqref{lemma_condi_a} or condition~\eqref{lemma_condi_b} holds. Then, we have

\begin{equation}
	L_{w}(\theta) = 2\pi \delta(\theta) + \sum_{u=1}^{p}\sum_{t=-\infty}^{\infty}d_{t}(u)e^{-iuw}e^{-it\theta} \nonumber
\end{equation}
for $-\pi< w,\theta \leq \pi$, where
\begin{equation}
	\left|\sum_{u=1}^{p}\sum_{t=-\infty}^{\infty}d_{t}(u)e^{-iuw}e^{-it\theta} \right|\leq \sum_{u=1}^{p}\sum_{t=-\infty}^{\infty}|d_{t}(u)|<\infty \nonumber.
\end{equation}

\smallskip
For any $-\pi<w,\theta\leq \pi$, we have $|L_{w}(\theta)|<\infty$ for $\theta \neq 0$ and $|L_{w}(0)| = \infty$. Thus the slowly varying condition is satisfied.

The examples are in fact more general than they seem to be. Specifically, the filter $d_{t}(u)$ in the first example can be generalized in different ways. For example, one can shift and scale the cosine function kernels with some constants, and modify the coefficient $1/2^{k}$. The zero-order term $d_{t}(0)=1$ for both examples can be replace by other function forms. To see this, one can multiply a function $a(t)$ with Fourier transform $A(w)$ to $d_{t}(u)$, then the function $L_{w}(\theta)$ is convolved with $A(w)$. If $|A(w)|$ is highly concentrated around the zero frequency, then the slowly varying condition on $L_{w}(\theta)$ could be preserved after the convolution. Meanwhile, since $a(t)d_{t}(u) = |A(0)|d_{t}(u)+(a(t)-|A(0)|)d_{t}(u)$, where the first term is dominating, the invertibility condition could hold for $a(t)d_{t}(u)$. 

\subsection{Stationarity test}
In order to verify the stationarity of the residual processes $\hat{N}_{t} \triangleq Y_{t}- \sum_{u=0}^{p} \hat{d}_{t}(u) X_{t-u}$ as in Algorithm~\ref{alg:1}, we make use of an improved version of the original Priestley and Subba Rao stationary test (PSR test)~\cite{priestley1969} by incorporating the multitaper method~\cite{xiang2019,constantine2011} to obtain $\{\hat{f}^{K}_{t_{i}}(w_{j}),1 \leq i \leq I, 1\leq j \leq J \}$ in Algorithm~\ref{alg:2}. The null hypothesis $H_{0}$  is ``$\{X_{t}\}$ is stationary". Consider a semi-stationary process $\{X_{t},0 \leq t \leq T\}$, let $f_{t}(w)$ denote its evolutionary spectral density and $\hat{f}^{K}_{t}(w)$ denote the multitaper estimate with $K$ tapers and $N$ as the length of the sample records. For $i \in \{1,\ldots,I\},$ with $I = \lfloor T/N \rfloor$,  and $j \in \{1,\ldots,J\}$, with $J =  \lfloor(N+1)/(K+1)\rfloor$, let $W_{ij} \triangleq \log\hat{f}^{K}_{t_{i}}(w_{j})+\psi(k)+\log K$, where $\psi(\cdot)$ is the digamma function.
The stationary test is based on applying the two-way analysis of variance (ANOVA) test to
 $\{W_{ij}\}$.
  Let 	$\textstyle W_{\cdot\cdot}=(1/IJ)\sum_{i=1}^I \sum_{j=1}^J W_{ij}, W_{i\cdot}=(1/J)\sum_{j=1}^J W_{ij}$, and $\textstyle W_{\cdot j}=(1/I)\sum_{i=1}^I W_{ij}$. The following statistics are needed: between time variance $S_T=J\sum_{i=1}^I (W_{i\cdot}-W_{\cdot\cdot})^2$; 
between frequencies variance $S_F=I\sum_{j=1}^J (W_{\cdot j}-W_{\cdot\cdot})^2$; interaction and residual variance
    $S_{I+R}=\sum_{i=1}^I \sum_{j=1}^J (W_{ij}- W_{i\cdot }-W_{\cdot j}+W_{\cdot\cdot})^2$. The algorithm is described in Algorithm~\ref{alg:2}, where testing $S_{I+R}/\s^2\sim\chi^2_{(I-1)(J-1)}$ is essentially a UMP test (see details from~\cite{priestley1969}).
\begin{algorithm}[H]

\makeatletter
% start with some helper code
% This is the vertical rule that is inserted
\newcommand*{\algrule}[1][\algorithmicindent]{%
  \makebox[#1][l]{%
    \hspace*{.2em}% <------------- This is where the rule starts from
    \vrule height 0.75\baselineskip depth 0.4 \baselineskip
  }
}

\newcount\ALG@printindent@tempcnta
\def\ALG@printindent{%
    \ifnum \theALG@nested>0% is there anything to print
    \ifx\ALG@text\ALG@x@notext% is this an end group without any text?
    % do nothing
    \else
    \unskip
    % draw a rule for each indent level
    \ALG@printindent@tempcnta=1
    \loop
    \algrule[\csname ALG@ind@\the\ALG@printindent@tempcnta\endcsname]%
    \advance \ALG@printindent@tempcnta 1
    \ifnum \ALG@printindent@tempcnta<\numexpr\theALG@nested+1\relax
    \repeat
    \fi
    \fi
}
% the following line injects our new indent handling code in place of the default spacing
\patchcmd{\ALG@doentity}{\noindent\hskip\ALG@tlm}{\ALG@printindent}{}{\errmessage{failed to patch}}
\patchcmd{\ALG@doentity}{\item[]\nointerlineskip}{}{}{} % no spurious vertical space
% end vertical rule patch for algorithmicx
\makeatother

   % \footnotesize
	    \caption{PSR stationarity test~\cite{priestley1969} \label{alg:2} }
	   
	\begin{algorithmic}
    	\State {\textbf{Input:} $\{\hat{f}^{K}_{t_{i}}(w_{j}),1 \leq i \leq I, 1\leq j \leq J \}$}
    	\State {\textbf{Output:} accept or reject $H_{0}$}
    		\State Compute $\{W_{ij},1 \leq i \leq I, 1\leq j \leq J \}$
    	%\State {\bf UMP test}:
        \State   Test  $S_{I+R}/\s^2\sim\chi^2_{(I-1)(J-1)}$
        \If {significant}
        \State reject $H_{0}$
        \Else
    	    \State Test $S_T/{\s^2}\sim\chi^2_{(I-1)}$
    	    \If {significant}
                 \State reject $H_{0}$
            \Else
                \State accept $H_{0}$
            \EndIf
    	\EndIf
	\end{algorithmic}
\end{algorithm}

\section{Extension to the network setting}
\label{sec:network}
 Our bivariate model shows how nonstationarity can be used for identifying the causal relation of a pair of processes. A more general setting is to identify the causal relations of a set of processes that corresponds to a DAG. In this section, we continue to exploit nonstationarity for identifying the underlying DAG of a set of Gaussian processes.

For a DAG $\mathcal{G}$ with nodes $\boldsymbol{V} = \{1,\ldots, N\}$, we use $\boldsymbol{PA}(j)$ and $\boldsymbol{ND}(j)$ to denote the set  of parents and set of non-descendents of a node $j \in \boldsymbol{V}$, respectively. The set of non-descendents $\boldsymbol{ND}(j)$ is the set of all nodes in $\boldsymbol{V}$ such that there is no path from $j$ to any $k \in \boldsymbol{ND}(j)$. 
\subsection{Model identifiability}

Consider a $N$ dimensional Gaussian  process $\{\boldsymbol{X}_{t}\} = \{\{X^{1}_{t}\}, \ldots,  \{X^{N}_{t}\}\}$ generated by the following SCM corresponding to a DAG $\mathcal{G}$ with nodes $\boldsymbol{V}=\{1,\ldots,N\}$, 
\begin{equation}
    X^{j}_{t} = \sum_{k\in \boldsymbol{PA}(j)}\Phi^{k \to j}_{t}(\mathsf{B})  X^{k}_{t}  + N^{j}_{t}, \quad j \in \boldsymbol{V}, \label{model_multi}
\end{equation}
where the noise processes $\{N^{j}_{t}\},j \in \boldsymbol{V}$, are jointly independent and stationary. The maximal order of the operators $\Phi^{k \to j}_{t}(\mathsf{B})$'s is $p$. 	
\smallskip

\begin{remark}
	Note that when $N=2$, model~\eqref{model_multi} reduces to a bivariate model with a stationary process as the cause. While in our bivariate model~\eqref{eq_forw}, the generating process of the cause $\{X_{t}\}$ is not specified. Thus model~\eqref{model_multi} is the extension of a special case of our bivariate model~\eqref{eq_forw} to the network setting. 
\end{remark}
\smallskip

% By applying an invertible operator $\Theta_{t}^{r}(\mathsf{B})$ on both sides of~\eqref{model_multi}, we obtain 
%\begin{equation}
%    \Theta_{t}^{r}(\mathsf{B})X^{j}_{t} = \sum_{k\in \boldsymbol{PA}(j)} \Theta_{t}^{r}(\mathsf{B})\Phi^{k \to j}_{t}(\mathsf{B})  X^{k}_{t}  + \widetilde{N}_{t}^{j}
%    . \label{model_multi3}
%\end{equation}
%If noise process $\{\widetilde{N}_{t}^{j}\}$ defined by $\widetilde{N}_{t}^{j} =  \Theta_{t}^{r}(\mathsf{B})N^{j}_{t} $ is stationary, then 
% The stationarity of $\{\widetilde{N}_{t}^{j}\}$ holds in general when $\Theta_{t}^{r}(\mathsf{B})$ is time-invariant. Thus we focus on the cases when there is no time-varying feedbacks from the past of $X_{t}^{j}$ (i.e., $ \Theta_{t}^{r}(\mathsf{B})$ is time-invariant). 

As a consequence of the time-varying operators in~\eqref{model_multi}, $\{\boldsymbol{X}_{t}\}$ is a set of stationary/nonstationary processes. Then a natural question is: What kind of nonstationarity is needed for identifying the DAG $\mathcal{G}$? First, let us start with the following example to show that the time-varying operators may not lead to nonstationarity.
 
\smallskip  

{\noindent \bf Example 6.}
Let $Y_{t} = \Phi^{2}_{t}(\mathsf{B})X_t$, where $\{X_{t}\}$ is \iid and $\Phi^{2}_{t}(\mathsf{B})$ is defined by $\phi_{t,0}=(-1)^{t}$ and  $\phi_{t,1}=(-1)^{t-1}$. Then $\{Y_{t}\}$ is stationary since $\E[Y_t] = (-1)^{t}\mu_{X} +(-1)^{t-1}\mu_{X} = 0$, $\Cov(Y_{t},Y_{s}) = 2\sigma_{X}^{2}$ for $t=s$, $\Cov(Y_{t},Y_{s}) = \sigma_{X}^{2}$ for $|t-s| = 1$, and $\Cov(Y_{t},Y_{s}) =0$ for $|t-s| \geq  2$. 
\smallskip  

This example can be easily generalized by applying any time-invariant operator to $Y_{t}$. Thus, simply using time-varying operators in~\eqref{model_multi} may not lead to nonstationarity in some non-generic cases. In order for the complete graph $\mathcal{G}$ to be identifiable, we need the following assumption.

\smallskip  
\begin{assumption}
A process $\{X^j_{t}\}, j\in \boldsymbol{V},$ remains nonstationary if one conditions on $\{X_{t}^{\boldsymbol{S}} =\boldsymbol{0} \}$, where $\boldsymbol{PA}(j) \not \subseteq \boldsymbol{S} \subseteq \boldsymbol{ND}(j)\setminus j$.
\label{assump1} 
\end{assumption}
\smallskip

In general, Assumption~\ref{assump1} is satisfied when all operators in~\eqref{model_multi} are time-varying and chosen generically. In the \iid setting, the restricted ANMs~\cite{peters2014causal} assume that the model $X_{j} = f_{j}(X_{\boldsymbol{PA}(j)})+N_{j}$ belongs to a bivariate identifiable class if one conditions on $X_{\boldsymbol{PA}(j)\setminus k} = \boldsymbol{x}$ for each $k \in \boldsymbol{PA}(j)$. In particular, for a model with Gaussian noise, the function $f_{j}$ needs to remain nonlinear when $X_{\boldsymbol{PA}(j)\setminus k }=\boldsymbol{x}$ are conditioned on, which is similar to how we require the nonstationarity to exist when $\{X_{t}^{k} = \boldsymbol{x_{\boldsymbol{S}}}\}$, $k \in  \boldsymbol{S}$, are conditioned on.

 Let $\boldsymbol{R} \subseteq \boldsymbol{V}$ denote the set of root nodes in $\mathcal{G}$ (i.e., all nodes $j$'s such that $\boldsymbol{PA}(j)= \varnothing$). The identifiability of the graph $\mathcal{G}$ is built on the following lemma and the definition of \emph{causal ordering}.

\begin{lemma}
A process $\{X_{t}^{j}\}, j \in \boldsymbol{V},$  is stationary if and only if  $j \in \boldsymbol{R}$. \label{lem1}
\end{lemma}

\begin{IEEEproof}
$\Longrightarrow:$ For any $j \in \boldsymbol{R}$, the process $\{X_{t}^{j}\}$ is determined by $X_{t}^{j} = N_{t}^{j}$ and thus it is stationary. $\Longleftarrow:$ If no process (i.e., the empty set) is conditioned on, then Assumption~\ref{assump1} implies that $\{X^{j}_{t}\}$ is nonstationary if $\boldsymbol{PA}(j) \neq\varnothing$ (i.e. $j \not \in \boldsymbol{R}$). Therefore, any process $\{X_{t}^{j}\}, j \in \boldsymbol{V},$  is stationary if and only if  $j \in \boldsymbol{R}$. 
\end{IEEEproof}

\begin{definition}[Causal ordering]
	A \emph{causal ordering} of the nodes $\boldsymbol{V}$ of a DAG $\mathcal{G}$ is an ordering of $\boldsymbol{V}$ such that there is no path from a later node to any earlier node.
\end{definition}
\smallskip

From the definition of causal ordering, the parents of each node in $\boldsymbol{V}$ are contained in the previous nodes, which motivates the proof the following theorem. 

\smallskip
\begin{theorem}
\label{thm2}
The graph $\mathcal{G}$ entailed in~\eqref{model_multi} is identifiable. 
\end{theorem}

\smallskip

\begin{IEEEproof}
First, we classify the nodes $\boldsymbol{V}$ to $K$ classes $\{\boldsymbol{V}^{1}, \ldots, \boldsymbol{V}^{K}\}$ as follows. Since the set of root nodes $\boldsymbol{R}$ is identifiable by Lemma~\ref{lem1}, let $\boldsymbol{V}^{1} = \boldsymbol{R}$. For $k \geq 2$, by conditioning on the processes $\{X_{t}^{j}\}, j \in \boldsymbol{V}^{i-1}, i \leq k$, to be zeros, we define $\boldsymbol{V}^{k}$ as the nodes in $\boldsymbol{V} \setminus \cup_{i=1}^{k-1} \boldsymbol{V}^{i}$ such that the corresponding processes are stationary. The iteration stops if $ \cup_{i=1}^{k} \boldsymbol{V}^{i} = \boldsymbol{V} $. The iteration will stop within $K \leq N$ steps due to the existence of a (unknown) causal ordering.

By Assumption~\ref{assump1}, the conditioning step implies that $\boldsymbol{PA}(j) \subseteq \cup_{i=1}^{k-1}\boldsymbol{V}^{i}$ for each $j \in \boldsymbol{V}^{k}$ and $2 \leq  k \leq K$, which means that the parents of each node are in previous classes. Again, Assumption~\ref{assump1} implies that $\boldsymbol{PA}(j)$ of $j \in \boldsymbol{V}^{k}$ is the smallest set $\boldsymbol{S} \subseteq \cup_{i=1}^{k-1}\boldsymbol{V}^{i}$ such that $\{X_{t}^{j}\}$ is stationary when the processes that correspond to $\boldsymbol{S}$ are conditioned on to be zeros. Since the parents of each node $j \in \boldsymbol{V}$ are identified, the graph $\mathcal{G}$ is identifiable.
 \end{IEEEproof}

 \subsection{Model estimation}

In Section~\ref{est_transf}, we described an estimation procedure of the time-varying filter for the bivariate model~\eqref{model_transfer}, while the estimation of time-varying filters for general multivariate models remains an open problem. Our causal inference procedure for the network setting is motivated by the following observation. By replacing each $X_{t}^{k}$ in~\eqref{model_multi} with the corresponding structural equation iteratively, we obtain an equivalent representation of model~\eqref{model_multi},  
\begin{equation}
    X^{j}_{t} = \sum_{k\in \boldsymbol{AN}(j)}\Psi^{k \to j}_{t}(\mathsf{B})  N^{k}_{t} + N_{t}^{j}, \quad j \in \boldsymbol{V}, \label{model_multi2}
\end{equation}
where $\boldsymbol{AN}(j)$ denotes the set of ancestors of the node $j$ (i.e., all nodes $k$'s such that there exists a path from $k$ to $j$) and each operator $\Psi^{k \to j}_{t}(\mathsf{B})$ is given by
\begin{equation}
	\Psi^{k \to j}_{t}(\mathsf{B}) = \sum_{(k,v_{1}, v_{2}, \ldots, j)}  \Phi^{k \to v_{1}}_{t}(\mathsf{B})  \Phi^{v_{1}\to v_{2}}_{t}(\mathsf{B}) \ldots \Phi^{v_{d} \to j}_{t}(\mathsf{B}),
	\nonumber
\end{equation}
where $(k,v_{1}, v_{2}, \ldots , j)$ denotes any path of any length $d+1$ from $k$ to $j$. Note that the operator $\Psi^{k \to j}_{t}(\mathsf{B})$ in~\eqref{model_multi2} and the operator $\Phi^{k \to j}_{t}(\mathsf{B})$ in~\eqref{model_multi} are equal for each $j$ and $k \in \boldsymbol{PA}(j)$. In our algorithm for the network setting, we use $d_{t}^{k \to j}(u)$ to denote $\Psi^{k \to j}_{t}(\mathsf{B})$ as in Algorithm~\ref{alg:1}. Since $X_{t}^{j}$ is written as a time-dependent linear combination of jointly independent variables, we estimate each filter in~\eqref{model_multi2} using the pairwise procedure described in Section~\ref{est_transf}, which turns out to perform well empirically. While $\{N_{t}^{k}\}$ is not observed if $k$ is not a root node, we will see later that our algorithm naturally provides estimates of the residuals. 
%{\color{red}The pairwise estimation problem remains to be challenging since the SNR is often less than $1$ for each pair. }
 
Based on model \eqref{model_multi2}, our algorithm first identifies the ancestors of a node $j$. Then the
 task is to identify the parents of $j$ given its ancestors. Implied by Assumption~\ref{assump1}, $\boldsymbol{PA}(j)$ is the smallest set $Q \subseteq\boldsymbol{AN}(j)$ such that $\{X_{t}^{j}\}$ is stationary when $\{X_{t}^{k}\}, k \in Q$, are conditioned on to be zeros. But such conditioning is hard to evaluate in practice. To introduce our procedure for identifying $\boldsymbol{PA}(j)$ (i.e., Procedure~\ref{pro:2}), we need the following assumption, which is again generally satisfied, based on which we show the correctness of Procedure~\ref{pro:2} in the proposition below.
 
\smallskip
\begin{assumption}\label{assum2}
	For any $Q \subseteq \boldsymbol{AN}(j)$ such that $Q \neq \boldsymbol{PA}(j)$, the equation
	\begin{equation}
	W_{t}^{j} =	X_{t}^{j} - \sum_{k\in Q}\Psi^{k \to j}_{t}(\mathsf{B})  X^{k}_{t} \label{assum2_eq}
	\end{equation}
determines a nonstationary process $\{W^{j}_{t}\}$.	
\end{assumption}
\smallskip

\begin{proposition}For any $Q \subseteq \boldsymbol{AN}(j)$, the process $\{W_{t}^{j}\}$ determined by~\eqref{assum2_eq} is stationary if and only if $Q = \boldsymbol{PA}(j)$
\end{proposition}

It is straightforward to see that we obtain $W_{t}^{j} = N_{t}^{j}$ when $Q=\boldsymbol{PA}(j)$ in~\eqref{assum2_eq}, using $\Psi^{k \to j}_{t}(\mathsf{B}) = \Phi^{k \to j}_{t}(\mathsf{B})$ and~\eqref{model_multi}. Thus $\{W_{t}^{j}\}$ is stationary. The other direction is a direct consequence of Assumption~\ref{assum2}.

\begin{algorithm}[H]

\makeatletter
% start with some helper code
% This is the vertical rule that is inserted
\newcommand*{\algrule}[1][\algorithmicindent]{%
  \makebox[#1][l]{%
    \hspace*{.2em}% <------------- This is where the rule starts from
    \vrule height 0.83\baselineskip depth 0.26 \baselineskip
  }
}

\newcount\ALG@printindent@tempcnta
\def\ALG@printindent{%
    \ifnum \theALG@nested>0% is there anything to print
    \ifx\ALG@text\ALG@x@notext% is this an end group without any text?
    % do nothing
    \else
    \unskip
    % draw a rule for each indent level
    \ALG@printindent@tempcnta=1
    \loop
    \algrule[\csname ALG@ind@\the\ALG@printindent@tempcnta\endcsname]%
    \advance \ALG@printindent@tempcnta 1
    \ifnum \ALG@printindent@tempcnta<\numexpr\theALG@nested+1\relax
    \repeat
    \fi
    \fi
}
% the following line injects our new indent handling code in place of the default spacing
\patchcmd{\ALG@doentity}{\noindent\hskip\ALG@tlm}{\ALG@printindent}{}{\errmessage{failed to patch}}
\patchcmd{\ALG@doentity}{\item[]\nointerlineskip}{}{}{} % no spurious vertical space
% end vertical rule patch for algorithmicx
\makeatother
    %\footnotesize
%        \captionsetup{font=footnotesize}
 %   \SetAlgoLined
 %    \DontPrintSemicolon
	    \caption{Causal inference procedure (network) \label{alg:3}
}
	\begin{algorithmic}
    	\State {\textbf{Input:} $N$ time series $\{x_{t}^{i}\}, \quad  i \in \boldsymbol{V}, \quad t =1 ,\ldots,T$}
    	\State {\textbf{Output:} adjacency matrix $A$,  \quad estimated residuals $\{\hat{N}_{t}^{i}\}$}
\State {\textbf{Initialization:}}  $S =  \boldsymbol{V} $, \enspace $A= \boldsymbol{0}_{N \times N}$, \enspace $\{\hat{N}_{t}^{i}\} = \{x_{t}^{i}\}$
\While{$S \neq \varnothing$}

\State $i^{*} = \text{MinStationary}(\{\hat{N}_{t}^{i}, i \in S\})$
\If {$i^{*}  = \varnothing$} {break} \EndIf
\State $S \leftarrow S \setminus i^{*}$
\For{$j \in S$}
\State {Initialization: }$c^{j}_{t} = \boldsymbol{0}_{T \times 1}, \quad \boldsymbol{AN}_{j} = \varnothing$
    \For{$k\in S^{c}$}
        \State Estimate the filter $d^{k \to j}_{t}(u)$
        \If { $\{x_{t}^{j}\} \not \independent \{\hat{N}_{t}^{k}\}$ }
        \State $\boldsymbol{AN}_{j} \leftarrow \boldsymbol{AN}_{j} \cup k$
          \State $c^{j}_{t} \leftarrow c^{j}_{t} + \textstyle\sum_{u}\hat{d}_{t}^{k \to j}(u)\hat{N}_{t-u}^{k} $
        \EndIf
      \EndFor
     \State $n_{t}^{j} = x_{t}^{j} - c^{j}_{t}$
\EndFor

\State $j^{*} = \text{MinStationary}(\{n_{t}^{j}, j \in S\})$
\If {$j^{*} \neq \varnothing$}
\State $\{\hat{N}_{t}^{j^{*}}\} = \{n_{t}^{j^{*}}\}$
\State $\boldsymbol{PA}_{j^{*}} = $\text{ SelectParents}$( \boldsymbol{AN}_{j^{*}}, \{x_{t}^{i}\} , \{\hat{d}_{t}^{k \to j^{*}}\})$
\State $A(i,j^{*}) =  1, \quad \forall i \in \boldsymbol{PA}_{j^{*}}$ 
\EndIf

 \EndWhile
	\end{algorithmic}
	
\end{algorithm}

Our algorithm follows the main idea of Theorem~\ref{thm2}. In each iteration of the while loop, the task is to identify the ancestors of one node in $S$ and then select the parents from the ancestors, where $S$ contains the nodes whose parents are unknown and $S^{c}$ denotes the complement of $S$. The order that the nodes leave the set $S$ is a causal ordering.  We will obtain an estimate of the residuals $\{N_{t}^{j}\}$ if the ancestors of $j$ are contained in $S^{c}$. Later, the estimated residuals will be used for the estimation of the filters. There are three places in the algorithm where we need to select the time series that minimizes some stationarity measure, which is carried out in  Procedure~\ref{pro:1}. Specifically, we use the UMP test (i.e., the interaction and residual variance $S_{I+R}$ in Algorithm~\ref{alg:2}) as a prescreening step and then compute the between time variance $S_{T}$ in Algorithm~\ref{alg:2} to quantify the stationarity of the time series.

For the independence test between $\{x_{t}^{j}\}$ and $\{\hat{N}_{t}^{k}\}$, one can use the kernel independence test for random processes~\cite{chwialkowski2014kernel}, which could be computationally demanding. An efficient approximation is to test whether $(1/ T) |\mu| =| \sum_{t} d_{t}^{k \to j}(u) |< a$ for $u=1, \ldots, q$. In practice, when our model assumptions are violated, one can test the joint independence of the estimated residuals $\{\hat{N}_{t}^{i}\}, i \in \boldsymbol{V}$, at the end of the algorithm. In Section~\ref{sec:exp},  this step is omitted since our algorithm is applied to the data generated by model~\eqref{model_multi}.

\begin{algorithm}[H]
\makeatletter
% start with some helper code
% This is the vertical rule that is inserted
\newcommand*{\algrule}[1][\algorithmicindent]{%
  \makebox[#1][l]{%
    \hspace*{.2em}% <------------- This is where the rule starts from
    \vrule height 0.83\baselineskip depth 0.26 \baselineskip
  }
}

\newcount\ALG@printindent@tempcnta
\def\ALG@printindent{%
    \ifnum \theALG@nested>0% is there anything to print
    \ifx\ALG@text\ALG@x@notext% is this an end group without any text?
    % do nothing
    \else
    \unskip
    % draw a rule for each indent level
    \ALG@printindent@tempcnta=1
    \loop
    \algrule[\csname ALG@ind@\the\ALG@printindent@tempcnta\endcsname]%
    \advance \ALG@printindent@tempcnta 1
    \ifnum \ALG@printindent@tempcnta<\numexpr\theALG@nested+1\relax
    \repeat
    \fi
    \fi
}
% the following line injects our new indent handling code in place of the default spacing
\patchcmd{\ALG@doentity}{\noindent\hskip\ALG@tlm}{\ALG@printindent}{}{\errmessage{failed to patch}}
\patchcmd{\ALG@doentity}{\item[]\nointerlineskip}{}{}{} % no spurious vertical space
% end vertical rule patch for algorithmicx
\makeatother
    %\footnotesize
%\captionsetup{font=footnotesize}
       \floatname{algorithm}{Procedure}

	    \caption*{\textbf{Procedure 1} MinStationary 
}
	\begin{algorithmic}
	    	\State {\textbf{Input:} $N$ time series $\{x_{t}^{j}\},\quad j \in J$  }\label{pro:1}
    	\State {\textbf{Output:} $ j^{*} \in J$ }
	\State $U = \{j \in J: \{x_{t}^{j}\} \text{ is a UMP}\}$	
	\If {$U \neq \varnothing$}
		\State $j^{*} = \textstyle\argmin_{j \in U}$  $S_{T}(\{x_{t}^{j}\})$ 
	\Else { $j^{*} =\varnothing$}
	\EndIf
	\end{algorithmic}
\end{algorithm}

\begin{algorithm}[H]
\makeatletter
% start with some helper code
% This is the vertical rule that is inserted
\newcommand*{\algrule}[1][\algorithmicindent]{%
  \makebox[#1][l]{%
    \hspace*{.2em}% <------------- This is where the rule starts from
    \vrule height 0.83\baselineskip depth 0.26 \baselineskip
  }
}

\newcount\ALG@printindent@tempcnta
\def\ALG@printindent{%
    \ifnum \theALG@nested>0% is there anything to print
    \ifx\ALG@text\ALG@x@notext% is this an end group without any text?
    % do nothing
    \else
    \unskip
    % draw a rule for each indent level
    \ALG@printindent@tempcnta=1
    \loop
    \algrule[\csname ALG@ind@\the\ALG@printindent@tempcnta\endcsname]%
    \advance \ALG@printindent@tempcnta 1
    \ifnum \ALG@printindent@tempcnta<\numexpr\theALG@nested+1\relax
    \repeat
    \fi
    \fi
}
% the following line injects our new indent handling code in place of the default spacing
\patchcmd{\ALG@doentity}{\noindent\hskip\ALG@tlm}{\ALG@printindent}{}{\errmessage{failed to patch}}
\patchcmd{\ALG@doentity}{\item[]\nointerlineskip}{}{}{} % no spurious vertical space
% end vertical rule patch for algorithmicx
\makeatother
       \floatname{algorithm}{Procedure}
	    \caption*{\textbf{Procedure 2} SelectParents}
 	\begin{algorithmic}
    	\State {\textbf{Input:} $\boldsymbol{AN}_{j}, \{x_{t}^{i}\}, \{\hat{d}_{t}^{k \to j}(u)\}$}
    	\State {\textbf{Output:} $\boldsymbol{PA}_{j}$}\label{pro:2}
	 \For {each $ Q \subseteq \boldsymbol{AN}_{j}$}	
		\State  $\hat{W}_{t}(Q) = x_{t}^{j} - \textstyle\sum_{ m \in Q}  \sum_{u}\hat{d}_{t}^{m \to j}(u)x_{t-u}^{k} $
	\EndFor
		\State $\boldsymbol{PA}_{j} = \text{MinStationary}(\{\hat{W}_{t}(Q)\})$
	\end{algorithmic}
\end{algorithm}

\begin{remark}
From our experiments on synthetic data (i.e., Experiment 5 in Section~\ref{sec:exp}), the selected parents $\boldsymbol{PA}_{j}$ in Procedure~\ref{pro:2} may be empty in certain cases, due to the estimation procedure. In such cases, one could replace $\boldsymbol{PA}_{j}$ with $\boldsymbol{AN}_{j}$ in Procedure~\ref{pro:2}, resulting in additional edges in the inferred graph (which is a subgraph of the transitive closure of $\mathcal{G}$~\cite{bang2008digraphs}). It is worth noting that the additional edges will not affect the causal ordering of nodes.
	\end{remark}

\section{Experiments}
\label{sec:exp}

For all data sets, we use $x$ and $y$ to denote the true cause and effect, respectively. For the independence test, we use the default configuration of  HSICp~\cite{chwialkowski2014kernel}. The significance level is denoted by $\a$ for the stationarity test, UMP test (i.e., the test on $S_{I+R}$ in algorithm~\ref{alg:2}) and the independence test, and we take $\a=0.05$ for the independence test throughout this section. For the multitaper method, finding the optimal window size is notoriously hard even for stationary processes. We thus set the window size $N_{F}$ to be $128$ for synthetic data, and the robustness of $N_{F}$ is tested in Experiment~1 as well as the real data simulations. In the synthetic experiments, since the true order $p$ is less than the maximum order $\lfloor N_F/2\rfloor$, i.e., the true model is in the model class, we adopt BIC for order  selection since it is consistent. For real data, we test both AIC and BIC. The length of the processes $N$ is fixed to $2048$ for all the synthetic data. We compare with TiMINo-linear~\cite{peters2013causal},  
TCM~\cite{huang2015identification}, LiNGAM-t~\cite{hyvarinen2010estimation} and Granger causality~\cite{granger1969investigating}.

%For the Multitaper method, the estimation of auto-/cross-spectral density at time $t$ is based on $\{X_{t},Y_{t}\}$ at the neighbors of $t$ (i.e. the time interval $[t-N_{F}/2,t+N_{F}/2]$). Since the start or the end of the time series does not fit a window of size $N_{F}$, the filter is estimated only for the interval $[N_{F}/2, N-N_{F}/2]$ (of length $N-N_{F}$). If $N$ is close to $N_{F}$, then the residuals (of length $N-N_{F}$) may be too short for independence test.  

\subsection{Synthetic Data\label{synth}}
%\subsection{First-order model}
%\label{exp:A}
\noindent{\bf Experiment 1: First-order Models.} We consider first-order models from~\cite{priestley1973},
\begin{equation}
    Y_{t} = X_{t} + a(t)X_{t-1}+N_{t},\quad 0 \leq t \leq T-1,
    \label{model_A}\end{equation}
where $\{X_{t}\}$ is a UMP defined by $X_{t} = b(t)Z_{t}$, with $b(t)$ being a Gaussian kernel $\mathcal{N}(\mu_{b},\sigma_{b}^{2})$. We choose $\mu_{b}=0.5T$ and $\sigma_{b}=0.2T$, with the same ratios to $T$ as in~\cite{priestley1973}. The process $\{Z_{t}\}$ is defined by a second-order AR model,  $Z_{t} = 0.8Z_{t-1}-0.4Z_{t-2} + \varepsilon_{t}$, in which $\{\varepsilon_{t}\}$ is a white Gaussian noise with $\varepsilon_{t} \sim \mathcal{N}(0,100^{2})$. The stationary noise process\footnote{
The only difference between the model in~\eqref{model_A} and that in~\cite{priestley1973} is that the latter considers $\{N_{t}\}$ to be a UMP process.} $\{N_{t}\}$ is defined by $N_{t} = 0.8N_{t-1}-0.16N_{t-2} + e_{t} $, where $\{e_{t}\}$ is a white Gaussian noise with $e_{t} \sim \mathcal{N}(0,\sigma_{N}^{2})$. 

\subsubsection{Different frequencies}
\begin{figure}
\vspace{-0.14in}
\centering
\includegraphics[width=.9\linewidth]{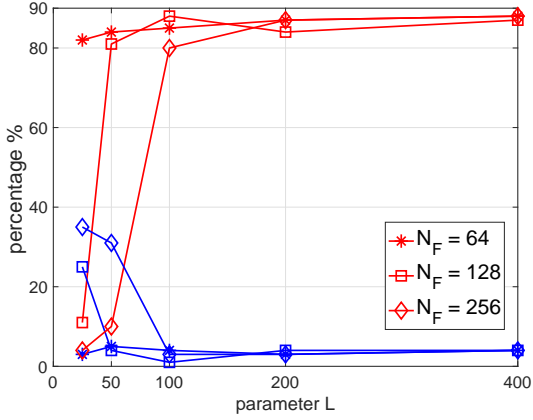}
\caption{\label{fig:a1} Synthetic data A1. The red lines denote $x \to y$, and the blue lines denote $y \to x$. For the clarity of the figure, we present the results for $N_{F}=64, 128,$ and $ 256$.} 
\label{fig1}
\vspace{-0.2in}
\end{figure}

We first test how the window size $N_{F}$ and the frequency of the cosine function $a(t)=0.5\cos(t/L)$ affect the performance of our method. Let $\sigma_{N}=25$, $L \in \{25,50,100,200,400\}$, and  $N_{F} \in \{32, 64, 128, 256, 512\}$. For each set of parameters, we test $100$ models. Fig.~\ref{fig1} shows that our method performs well for cosine functions with low frequencies ($L \geq 100$) regardless of the choice of $N_{F}$. For high frequencies (i.e., when $L$ is small), our method performs well only when $N_{F}$ is small. This aligns with the intuition that small $N_{F}$ can help reveal more high-frequency components. 

\subsubsection{Different SNRs}

\begin{figure}
\captionsetup[subfloat]{farskip=1pt,captionskip=-8pt}
\subfloat[\label{fig2a}]{\includegraphics[width=2.8cm]{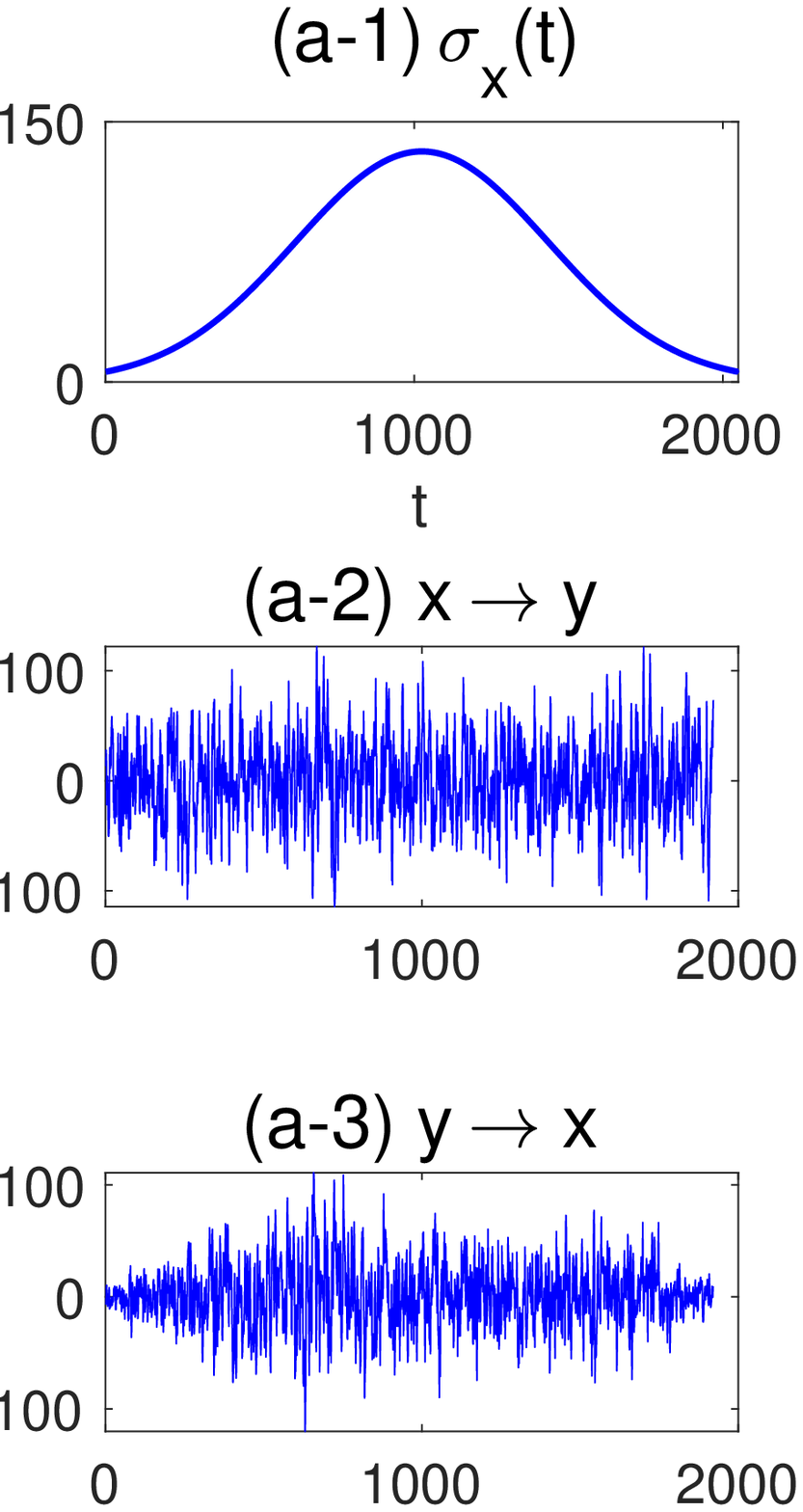}}\hspace{-0.45em}
\subfloat[\label{fig2b}]{\includegraphics[width=6.7cm]{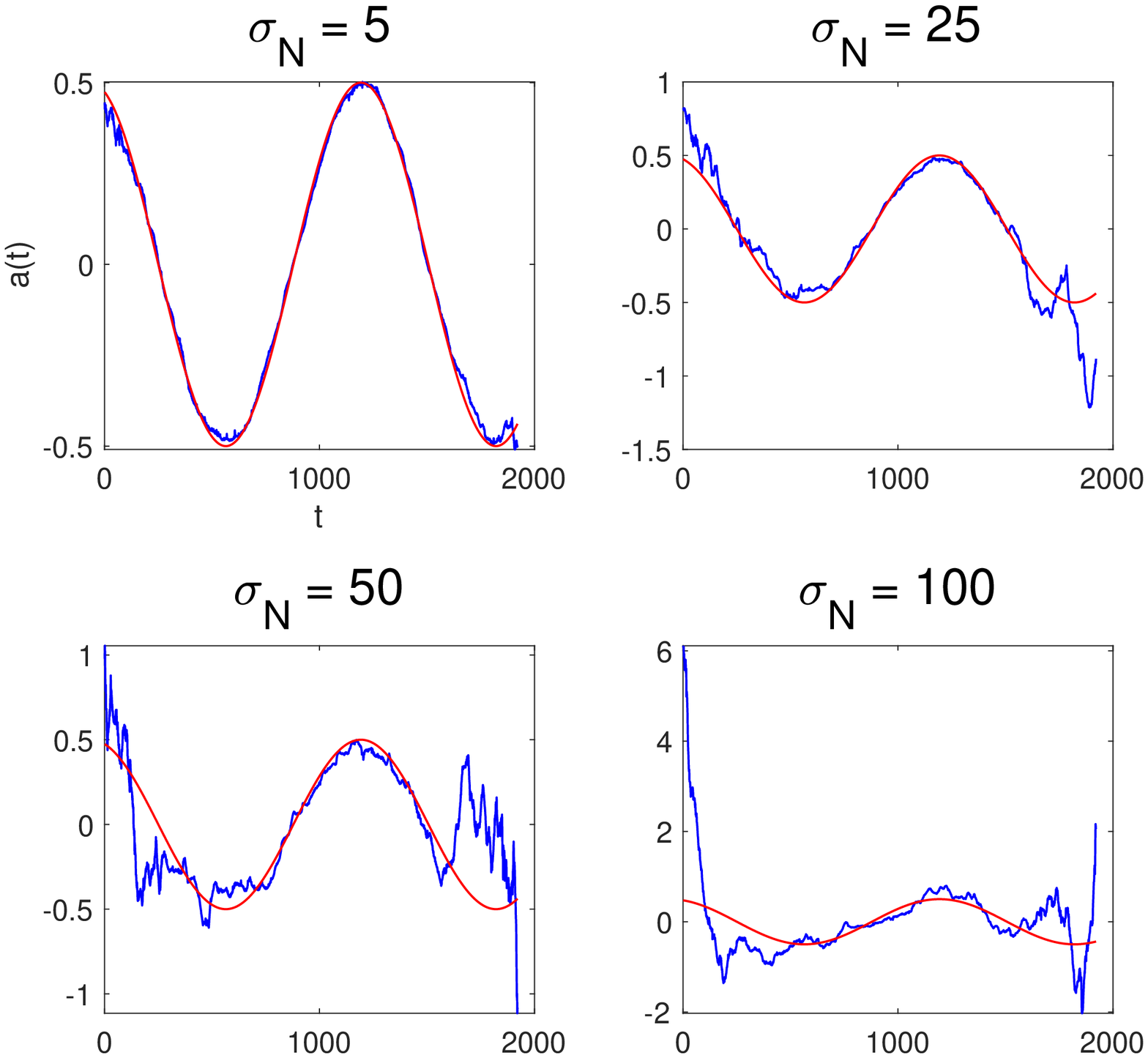}}
\hfill
\caption{
Synthetic data A2. (a-1) The standard deviation of $\{X_{t}\}$. (a-2) The estimated residuals of $x \to y$ when $\sigma_{N}= 25$.  (a-3) The estimated residuals of $y \to x$ when $\sigma_{N}= 25$. (b) Estimated $a(t)$'s for different $\sigma_{N}$. The red lines denote the ground truth, and the blue lines denote the estimate.}
\vspace{-0.2in}
\end{figure}

We now examine how sensitive our method is with respect to the SNR level. We use the parameter $\sigma_{N}$ to control the SNR level. Let $a(t) = 0.5 \cos(t/200)$. For each $\sigma_{N}$ in $\{5, 10, 15, 20, 25, 40, 55, 70, 85, 100\}$, we test $100$ models.
\label{exp:A2}
 For $\sigma_{N}=25$, one can tell from Fig. \ref{fig2a} that the residuals of $x \to y$ is more likely to be nonstationary than the residuals of $y \to x$. Overall, the percentage of identifying the correct directions is above $80\%$ for different SNRs. Though, Fig. \ref{fig2b} shows that the estimated cosine functions are noisier when the SNR is lower (i.e., when $\sigma_{N}$ is larger). Note that even for a fixed $\sigma_{N}$, the SNR changes over time with $\sigma_{X}(t)$ (Fig. \ref{fig2a}). So the estimated functions are noisier at the start and end of the time range. This suggests that our method is relatively robust with respect to different SNR levels.

%\subsection{High-order models}
%\label{exp:B}

\smallskip
\noindent{\bf Experiment 2: High-order Models.} In this experiment, we focus on high-order models with smooth filters. We generalize model $\eqref{model_A}$ to incorporate higher orders $p$ which is generate from $ \U \{1,\ldots,5\}$. Specifically, we have 
\begin{equation}
    Y_{t} = c(t) \sum_{k=0}^{p} a_{k}(t)X_{t-k}+N_{t},\quad 0 \leq t \leq T-1,
    \label{high_order}
\end{equation}
where $c(t) = 1+A \cos(t/L)$, $A \sim \U[0.05,0.2]$, $L \sim \U[400,800]$, $a_{0}(t)=1$, $a_{k}(t) = g_{k}(f(P(t/T)) + S(t))$, $k \geq 1$, in which $S(t) = \sinc((t-a)/b)$, $\sinc (t) \triangleq \sin(\pi t) / (\pi t)$, $a \sim \U[300,1500]$, $b \sim \U[400,800]$, and $P(t)$ is a polynomial function of degree $q \sim \U\{1,\ldots,6\}$ with roots sampled from $\U[-1,1]$. The functions $f$ and $g_{k}$'s are scaling functions defined as $\textstyle f(P(t)) \triangleq P(t)/ \max_{t}(|P(t)|)$ and $ \textstyle g_{k}(a_{k}(t)) \triangleq c_{k} a_{k}(t)/ \max_{t}(|a_{k}(t)|)$, with $c_{k}=(1/1.5)^{k}$. The process $\{X_{t}\}$ is defined by $X_{t}=Z_{t}+\sum_{k=1}^{2}b_{k}(t)Z_{t-k}$, where $b_{k}(t)$'s are generated by the same way as $a_{k}(t)$'s. The stationary processes $\{Z_{t}\}$ and $\{N_{t}\}$ are each generated by the ARMA model, $W_{t} = \Phi^{2}(\mathsf{B})W_{t}+ \Theta^{2}(\mathsf{B})e_{t}$, where $\Phi^{2}(\mathsf{B}) = (d_{1}+d_{2})\mathsf{B} - d_{1}d_{2} \mathsf{B}^{2}$, $\Theta^{2}(\mathsf{B}) = 1+ (d_{3}+d_{4})\mathsf{B} + d_{3}d_{4}\mathsf{B}^{2} $, and $d_{1},d_{2},d_{3},d_{4} \sim \U [-0.6,-0.1]\cup[0.1,0.6]$. Note that $1/d_{1}$ and $1/d_{2}$ are the roots of the polynomial $1-\Phi^{2}(z)$, and $1/d_{3}$ and $1/d_{4}$ are the roots of the polynomial $\Theta^{2}(z)$. Since the roots are all strictly outside the unit circle, the randomly generated ARMA model has a unique stationary solution that is causal~\cite{davis1987time}. The white Gaussian noise $\{e_{t}\}$ has variances $\sigma_{Z}^{2} = 100^2$ and $\sigma_{N}^{2}=25^{2}$ for $\{Z_{t}\}$ and $\{N_{t}\}$, respectively. For $\a = 0.01$ and $\a=0.05$, we test $1000$ randomly generated models. 

\begin{comment}
 To test the performance of our method for the models with lagged terms of $Y_{t}$ with time-invariant coefficients, we test $1000$ following models 
\begin{equation*}
    Y_{t} = b_{1}Y_{t-1} +b_{2}Y_{t-2} + X_{t} + \sum_{k=1}^{p} a_{k}(t)X_{t-k}+N_{t},
\end{equation*}
where $b_{1}$ and $b_{2}$ are generated in the same way as $d_{1}$ and $d_{2}$.
\end{comment}

In Table~\ref{tab:dataB}, we classify the undecided cases into three categories: (1) The independence tests are significant for both directions; (2) The independence tests are not significant and the residual processes are stationary; (3) The independence tests are not significant and the residual processes are nonstationary. Both TiMINo~\cite{peters2013causal} and TCM~\cite{huang2015identification} remain undecided since the p-values are too small for both directions. Granger causality infers the correct (or wrong) direction for $41\%$ (or $2.8 \%$) of the models. LiNGAM-t infers the correct (or wrong) direction for $69\%$ (or $31\%$) of the models.
 
\begin{table}[ht]
\normalsize
\centering
\caption{Experiment~2 results (\%)} 
\label{tab:dataB} 
\begin{tabular}{c|c|c}
\hline
 $\a$ for the stationarity test  & 0.01 &0.05  \\
\hline
 $x \to y$ & 84.1  & 82.8\\
 \hline
 $y \to x$ & 0.6 &0.9\\
 \hline
 both not independent & 11.7  & 10.8\\
   \hline
 both stationary & 1.0 &0.4\\
 \hline
  both nonstationary  & 2.6  & 5.1 \\
\hline
\end{tabular}
\end{table}

%The undecided cases fall into the the first category more often, which may due to the fail of the estimation procedure for certain filters.
% In practice, we suggest the use of 0.05.  Our results may not be able to show the weakness of the independence test or the stationarity test, since if the quality of the estimate of the filter is poor, it may lead to undecided cases.

\noindent{\bf Experiment 3: High-order Models with Gaussian/non-Gaussian UMP noise.} We replace the noise process in Experiment 2 with a UMP defined as $N_{t}=g(t)W_{t}$, where $g(t) = \exp((t-T/2)^{2}/(2\sigma^{2}_{g}))$, $\sigma_{g}\sim \U [0.4T, 0.8T]$. $\{W_{t}\}$ is simulated from the following three models: (1) the randomly generated ARMA model in Experiment 2 (i.e., $\{Z_{t}\}$) with $\sigma_{e} = 25$; (2) \iid uniform with $W_{t} \sim \U [-30,30]$; (3) $W_{t}= 15V_{t}$, with $\{V_{t}\}$ being a sequence of \iid variables following the student's t-distribution with degrees of freedom $5$. We use of a variant of Algorithm~\ref{alg:1} that replaces the stationarity test by a UMP test. We sample 1000 models from \eqref{high_order} for each case of the UMP noise. Our method works well for the three cases (see Table~\ref{tab:exp3}). In particular, the percentage of inferring the wrong direction is below $1\%$. LiNGAM-t tends to infer more wrong directions and Granger causality performs worse than a random guess (i.e., $50\%$).  TCM and TiMINo-linear remain mostly undecided.

\begin{table}[h!]
\small
\centering
\caption{Experiment~3 results (\%)} 
\label{tab:exp3} 
\begin{tabular}{c|c|c|c}
\hline
\bf{ Gaussian}  & Ours & LiNGAM-t & Granger\\
\hline
 $x \to y$ & 84.1  & 64.4  & 39.2\\
 \hline
 $y \to x$ & 0.7 & 35.6  & 2.5 \\
 \hline
undecided & 15.2  & 0  & 58.3  \\
   \hline \hline
\bf{uniform}    \\
\hline
 $x \to y$ & 84.2  & 66.1  & 29.9\\
 \hline
 $y \to x$ & 0.7 & 33.9  & 4.4 \\
 \hline
undecided & 15.1  & 0  & 65.7  \\
\hline \hline
\bf{student's t}      \\
 \hline
 $x \to y$ & 80.8  & 70.7  & 30.9\\
 \hline
 $y \to x$ & 0.3 & 29.3  & 3.5 \\
 \hline
undecided & 18.9  & 0  & 65.6  \\
\hline
\end{tabular}
\end{table}

\smallskip
\noindent{\bf Experiment 4: Models with Non-smooth Functions.} We have demonstrated the performance of our method for a large class of smooth filters in Experiment~2,   and we now examine a class of non-smooth functions. Consider model~\eqref{model_A} with $a(t)=0.5\tri(t/200;b), 0 \leq b < 1 $, which is a triangle wave function. The function $\tri(t;b)$ in one of its period $[0, 2\pi]$ is defined as
\begin{equation*}
     \tri(t;b) = 
\begin{cases}
    1-\frac{t}{2\pi},   &   0\leq t \leq 2\pi , b=0\\
    \frac{t}{2\pi b},   & 0\leq t \leq 2\pi b,  b\neq0\\
    \frac{t}{2\pi b- 2\pi }+\frac{1}{1-b},   & 2\pi b< t \leq 2\pi,  b\neq0
\end{cases},   
\end{equation*}
where $(2 \pi b,1)$ is a vertex of the triangle function that changes with $b$. The function $\tri(t;b)$ is a right triangle when $b=0$, which leads to a discontinuous point at $t=0$. Let $\sigma_{N}=25$. For each $b$ in $\{0,0.125,0.25,0.375,0.5\}$, we test 100 models. Table \ref{tab:dataC} shows that our method performs well except for the case when the triangle wave function has discontinuous points. 

\begin{table}[ht]
\normalsize
\centering
\caption{Experiment~4 results (\%)} 
\label{tab:dataC} 
\begin{tabular}{c|c|c}
\hline
  b & $x \to y$ & $y \to x$  \\
\hline
 0 & 33  & 4\\
 \hline
 0.125 & 89 &2\\
 \hline
   0.25 & 86  & 2\\
   \hline
 0.375 & 86 & 2\\
 \hline
  0.5  & 87  & 3\\
\hline
\end{tabular}
\end{table}

\smallskip
\noindent{\bf Experiment 5: Network Setting. \label{exp_net}} We consider randomly generated DAGs with number of nodes $N_{G} \sim \U \{2,\ldots, 5\}$. Each edge is included with probability $0.6$. In model~\eqref{model_multi2}, let $\{N^{j}_{t}\}$ be an \iid Gaussian process with zero mean and variance $\sigma^{2}_{j} \sim \U [5,10]$. Each time-varying filter is defined in the same way as the bivariate model~\eqref{high_order}, where $a_{k}(t)$ has order $2$, $c(t)=1+A \cos(t/L)$, with $A \sim \U [0.5,2]$ and $L\sim \U[400,800]$. We test Algorithm~\ref{alg:3} with $\alpha=0.01$ and $\alpha=0.05$ for the UMP test, respectively. To approximately test the independence between 
$\{x_{t}^{j}\}$ and $\{\hat{N}_{t}^{k}\}$ in Algorithm~\ref{alg:3}, we test whether $|\mu| = (1/ T) |\sum_{t} d_{t}^{k \to j}(u)| < a$, $a =0.15$,  for $u=1, \ldots, 3$. We test the sensitivity of our method with respect to parameter $a$ for $a =\{0.1, 0.125, 0.15, 1.75\}$ when $\alpha=0.05$. 

% To determine the parameter $a$,  we simulate two independent \iid standard Gaussian processes for $1000$ realizations. The length of each realization is $2048$.  For the $|\mu|$'s computed for the 1000 realizations, we taking the maximum of $|\mu|$ as the $a$.

%The results are classified into four categories: (1) the inferred graph is correct; (2) the inferred graph is the correct graph with additional edges and the causal order is correct; (3) the inferred graph is the correct graph with additional edges and the causal order is wrong; (4) all other cases (i.e.,  correct graph with missing edges or correct graph with both missing edges and additional edges). Our method mostly infers either the correct graph or the correct graph with additional edges while the causal order is correct. Both Granger causality and LiNGAM-t give a large percent of wrong graphs that include missing edges. TiMINo remains mostly undecided. 

The results are classified into three categories: (1) the
inferred graph is correct; (2) the inferred graph is a proper
subgraph of the correct graph (i.e., correct graph with missing
edges); (3) all other cases (i.e., correct graph with additional
edges or correct graph with both missing edges and additional
edges). We refer to the proper subgraph as p-subgraph in
Table~\ref{table:exp5}. When a proper subgraph is inferred, it means that our method tends to remain conservative. Our method mostly infers either the correct graph or
a proper subgraph of the correct graph. Both Granger causality
and LiNGAM-t give a large percent of wrong graphs that
include wrong edges. TiMINo remains mostly undecided. Overall, our method is relatively robust with respect to $a$ since the results mainly fall into the first two categories (see Table~\ref{table:exp5_2}). Our method infers more cases of wrong edges as $a$ gets smaller, thus we suggest using $a \geq 0.1$.

\begin{table}[ht]
\normalsize
\centering
\caption{Experiment 5 results (\%) \label{table:exp5}} 
\begin{tabular}{c|c|cc|c}
\hline
           & \multicolumn{2}{c|}{Ours}        & \multirow{2}{*}{Granger} & \multirow{2}{*}{LiNGAM-t} \\ \cline{2-3}
           & 0.01 & \multicolumn{1}{c|}{0.05} &                          &                           \\ \hline
correct graph  & 91.6 & \multicolumn{1}{c|}{81.7} & 32.4                     & 13.0                      \\ \hline
p-subgraph & 5.4  & \multicolumn{1}{c|}{14.2} & 14.1                     & 9.5                     \\ \hline
others     & 3.0  & \multicolumn{1}{c|}{4.1}  & 53.5                     & 77.5                      \\ \hline
\end{tabular}
\end{table}

\begin{table}[ht]
\normalsize
\centering
\caption{Sensitivity test of $a$ (\%) \label{table:exp5_2}} 
\begin{tabular}{c|c|c|c|c}
 \hline
$a$                     & 0.1 & 0.125 & 0.15 & 0.175 \\ \hline
correct graph         &    74.7     &    77.2  &  81.7    &   82.6   \\ \hline
p-subgraph &     14.2   & 16.5   &   14.2   &   13.6    \\ \hline
others                  &  11.1    &  6.3   &   4.1     &    3.8   \\ \hline
\end{tabular}
\end{table}

\subsection{Real Data}

Let $\a = 0.05$. The only preprocessing needed for our method is detrending. Since the first two data sets are too short in length for HSICp, we infer the causal directions using HSIC. In all experiments, we fix $N_{F} = 64$ but the p-values remain to be similar for $N_{F} = 32$ or $128$. It turns out that the p-values are similar under AIC or BIC. We set the maximum order for the compared methods (i.e., TiMINo, TCM, LiNGAM-t, Granger causality) to be 10. In all three experiments below, TCM~\cite{huang2015identification} remains undecided. 

%{\color{red}We perform tests with window sizes $N_{F}=\{32, 64,128\}$ for all data sets, and we conclude the causal directions with $N_{F}=128$. }
%
% Since the first two data sets are too short for the stationarity test and HSICp, we infer the causal directions using HSIC.}

%\subsection{Input Gas Rate vs. Output CO\textsubscript{2}}
\smallskip
\noindent{\bf Experiment 6: Input Gas Rate vs. Output CO\textsubscript{2}.} The Gas Furnace data set~\cite{box2015time} has two variables: the input gas rate ($x$) and the output CO$_{2}$ ($y$).  Our method with  HSIC yields the correct direction with p-values $p_{x \to y} = 0.0895$ and $p_{y \to x} = 5.511 \cdot 10^{-11}$. TiMINo and Granger causality correctly infer the direction $x \to y$~\cite{peters2013causal}, whereas LiNGAM-t gives the wrong direction.

%, while TCM is undecided since both p-values are bigger than $0.05$. 
 
%\subsection{Duration vs. Time Interval} 
\smallskip
\noindent{\bf Experiment 7: Duration vs. Time Interval.} The Old Faithful data set~\cite{azzalini1990look, Dua:2019} contains two variables observed from the Old Faithful geyser: the duration of an eruption ($x$) and the time interval before the next eruption ($y$). As mentioned in~\cite{peters2013causal}, the data is not collected with fixed time resolution, but we treat the data as time series. 
%This data set is available at~\cite{Dua:2019}.  
Our method equipped with HSIC leads to p-values $p_{x \to y} = 0.0777 $ and $p_{y \to x} = 9.364 \cdot 10^{-10}$. TiMINo and LiNGAM-t infer the correct direction~\cite{peters2013causal}. Granger causality gives the wrong direction.

%, while TCM is undecided since both p-values are small.

% \subsection{Breathing Pattern vs. Heart Rate}
\smallskip
\noindent{\bf Experiment 8: Breathing Pattern vs. Heart Rate.} The modulation of heart rate by the breathing pattern is known as the respiratory sinus arrhythmia (RSA)~\cite{hirsch1981respiratory}. This phenomenon is widely observed, especially among young and healthy individuals. We use the data from~\cite{rigney1994multi}\nocite{ichimaru1999development} to verify the causal relationship between breathing pattern and heart rate. Specifically, this data set contains two variables: chest volume ($x$) and heart rate ($y$). Since the mechanism of respiratory sinus arrhythmia is understood to be the synchronization of heart rate to breathing rhythm~\cite{hayano1996respiratory}, we consider $x$ as a cause for $y$. The challenging parts of the data are the nonstationarity and the seasonality of the data. Taking a segment of length $2048$ from the data (i.e., samples $2000\sim 4047$ of data set B1), our method yields p-values $p_{x\to y} = 0.0610$ and $p_{y\to x} = 0.004$. Since the stationarity tests are always significant, we conclude the causal directions based only on HSICp. The results show that our method gives the correct direction. TiMINo remains undecided due to small p-values ($<10^{-10}$) for both directions, and this might because it requires the stationarity of the data for the estimation procedure. Granger causality infers the correct direction, while LiNGAM-t gives the wrong direction.

\section{Acknowledgement}The authors would like to thank Jie Ding and Kun Zhang for inspiring discussions.

%\newpage

\appendices

\section{Proof of Theorem~\ref{thm:identification}}
\label{sec:thm1}
\begin{IEEEproof}
Suppose there exists a backward model of the form
\begin{equation}
        X_{t} =  \Psi_{t}^{q}(\mathsf{B})Y_{t} + \widetilde{N}_{t}, \quad \{\widetilde{N}_{t}\} \independent \{Y_{t}\}, \ \ q \in \bar{\mathbb{Z}}_{\geq 0},
        \label{pf_bck}
\end{equation}
where $ \{\widetilde{N}_{t}\}$ is stationary. 

%When $q=\infty$, by Proposition~\ref{prop:infty}, the series $\Psi_{t}^{\infty}(\mathsf{B})Y_{t}$ converges absolutely with probability one since $\E[Y_{t}]=0$. Furthermore, we have $\E\left[|Y_{t}|^{2}\right]<\infty$ due to the stationarity of $\{N_{t}\}$ (which implies that $\E[|N_{t}|^{2}] <\infty$~\cite{davis1987time}) and our assumption that $\E[|X_{t}|^{2}] <\infty$. Thus by Proposition~\ref{prop:infty}, the series $\Psi_{t}^{\infty}(\mathsf{B})Y_{t}$ converges absolute with probability one and converges in mean square to the same limit.

(I) First, we show the condition for the independence constraint. Let $\mathcal{O}$ denote the class of operators such that $\{Y_{t}\}\independent \{X_{t}-H^{s}_{t}(\mathsf{B})Y_{t}\}$ for any $H^{s}_{t}(\mathsf{B}) \in \mathcal{O}$, $s \in \bar{\mathbb{Z}}_{\geq 0}$. Note that $\mathcal{O}$ is nonempty since $\Psi_{t}^{q}(\mathsf{B}) \in \mathcal{O}$. For any $H^{s}_{t}(\mathsf{B}) \in \mathcal{O}$, we obtain the model
 \begin{equation}
 	X_{t} =  H^{s}_{t}(\mathsf{B})Y_{t} + W_t,
 	\label{back_Hs}
 \end{equation}
 where $\{W_{t}\}$ is defined by $W_{t} = X_{t}-H^{s}_{t}(\mathsf{B})Y_{t}$. We replace $Y_{t}$ in~\eqref{back_Hs} by that in~\eqref{defYt}, and have
\begin{align}
    W_{t} &= (1-H_{t}^{s}(\mathsf{B}) \Phi_{t}^{p}(\mathsf{B}))X_{t} - H_{t}^{s}(\mathsf{B})N_{t}.
    \label{ntilde}
\end{align}

 Now, we prove that the Gaussian process $\{H^{s}_{t}(\mathsf{B})Y_{t}\}$ has the same distribution for any $H^{s}_{t}(\mathsf{B}) \in \mathcal{O}$, which is equivalent to prove that $\{H^{s}_{t}(\mathsf{B})Y_{t}\}$ and $\{\Psi^{q}_{t}(\mathsf{B})Y_{t}\}$ have the same covariance function. According to the independence constraint, we know that both $\{W_t\}$ and $\{\widetilde{N}_t\}$ are independent of $\{Y_{t}\}$, which implies that $\{W_t-\widetilde{N}_t\}\independent \{Y_{t}\}$. 	Then, subtracting \eqref{pf_bck} from \eqref{back_Hs} yields $W_t -\widetilde{N}_t = (\Psi_{t}^{q}(\mathsf{B})-H^{s}_{t}(\mathsf{B}))Y_t$. It follows from Lemma~\ref{lem:cov} that the covariance function of $\{(\Psi_{t}^{q}(\mathsf{B})-H^{s}_{t}(\mathsf{B}))Y_t\}$ is zero. Thus, we have shown that $\{H_{t}^{s}(\mathsf{B})Y_{t}\}$ has the same distribution for any $H_{t}^{s}(\mathsf{B}) \in \mathcal{O}$.  
 	 
Moreover, we provide an explicit characterization of the operators in $\mathcal{O}$ as follows. Since two Gaussian processes are independent if and only if their cross-covariance function equals to zero, from~\eqref{defYt} and \eqref{ntilde}, we have 
\begin{align}
        &\Cov(Y_{t_{1}},W_{t_{2}})\nonumber\\
        &\hspace{2em}= \Cov(\Phi_{t_{1}}^{p}(\mathsf{B})X_{t_{1}} +N_{t_{1}}, \nonumber\\
        &\hspace{5em}(1-H_{t_{2}}^{s}(\mathsf{B}) \Phi_{t_{2}}^{p}(\mathsf{B}))X_{t_{2}} - H_{t_{2}}^{s}(\mathsf{B})N_{t_{2}})\nonumber\\
        &\hspace{2em}=  \Phi_{t_{1}}^{p}(\mathsf{B})(1-H_{t_{2}}^{s}(\mathsf{B}) \Phi_{t_{2}}^{p}(\mathsf{B}))\gamma_{XX}(t_{1},t_{2}) \nonumber\\
        & \hspace{10em}-H_{t_{2}}^{s}(\mathsf{B})\gamma_{NN}(t_{2}-t_{1})=0. \label{eq:cov}
\end{align}
Since we assume $\Phi_{t}^{p}(\mathsf{B})$ is invertible, \eqref{eq:cov} can be written as 
\begin{align}
\label{two_station}
   &(1-H_{t_{2}}^{s}(\mathsf{B}) \Phi_{t_{2}}^{p}(\mathsf{B}))\gamma_{XX}(t_{1},t_{2}) \nonumber\\
   & \hspace{7em}=H_{t_{2}}^{s}(\mathsf{B})  (\Phi_{t_{1}}^{p}(\mathsf{B}))^{-1}\gamma_{NN}(t_{2}-t_{1}),
\end{align}
which can be further simplified as 
\begin{align}
    \gamma_{XX}(t_{1},t_{2}) &= H_{t_{2}}^{s}(\mathsf{B})( \Phi_{t_{2}}^{p}(\mathsf{B})\gamma_{XX}(t_{1},t_{2})\nonumber\\
    &\hspace{5em} + (\Phi_{t_{1}}^{p}(\mathsf{B}))^{-1}\gamma_{NN}(t_{2}-t_{1}))
    \label{eq_solve_psi}.
\end{align}
Therefore we have shown that~\eqref{eq_solve_psi} determines a class of operators in $\mathcal{O}$, simply because that~\eqref{eq_solve_psi} is equivalent to the independence of $\{Y_{t}\}$ and $\{W_{t}\}$. 

(II) Now we move on to prove the condition for the stationarity constraint. Since when $H_{t}^{s}(\mathsf{B}) = \Psi_{t}^{q}(\mathsf{B})$, we obtain $\{W_{t}\}=\{\widetilde{N}_{t}\}$ in \eqref{back_Hs}. Thus there exists $H_{t}^{s}(\mathsf{B})\in \mathcal{O}$ such that $\{W_{t}\}=\{X_{t}-H_{t}^{s}(\mathsf{B})Y_{t}\}$ is stationary. The stationarity of $\{W_{t}\}$ implies that its covariance function is a function of $t_{2}-t_{1}$. From~\eqref{ntilde} and \eqref{two_station}, we have
\begin{align}
     &\Cov(W_{t_{1}},W_{t_{2}})\nonumber\\
                &\hspace{.5em}=    (1-H_{t_{1}}^{s}(\mathsf{B}) \Phi_{t_{1}}^{p}(\mathsf{B}))(1-H_{t_{2}}^{s}(\mathsf{B}) \Phi_{t_{2}}^{p}(\mathsf{B}))\gamma_{XX}(t_{1},t_{2})\nonumber\\
            &\hspace{12em}+H_{t_{1}}^{s}(\mathsf{B}) H_{t_{2}}^{s}(\mathsf{B})\gamma_{NN}(t_{2}-t_{1})\nonumber\\
                &\hspace{.5em}= \    H_{t_{2}}^{s}(\mathsf{B}) (\Phi_{t_{1}}^{p}(\mathsf{B}))^{-1}\gamma_{NN}(t_{2}-t_{1})\label{last_eq}.
\end{align}
Let $t_{1}=t_{2}=t$ in \eqref{last_eq} and  $\Theta_{t}^{r}(\mathsf{B}) \triangleq  (\Phi_{t}^{p}(\mathsf{B}))^{-1}$, we obtain the variance of $\{W_{t}\}$ as 
\begin{equation}
 \Var(W_{t})= \sum_{j=1}^{s}\sum_{k=1}^{r}\eta_{t,j}\theta_{t-j,k}\gamma_{NN}(k-j),
 \label{var_nt}
\end{equation}
which is time-invariant. 
\end{IEEEproof}

\section{Proof of Corollary~\ref{coro:iidsetting}}
\label{sec:coro1}

\begin{IEEEproof} 
Note that $\E[X_{t}X_{s}]=\sigma^{2}_{X}\mathds{1}_{t=s}$ since $\{X_{t}\}$ is an \iid process. We solve for the coefficients of $\Psi^{q}_{t}(\mathsf{B})$ by computing $\E[X_{t}Y_{t+p-i}]$, $i\geq 0$. Using model \eqref{defYt}, we derive	
	\begin{align*}
		\E[X_{t}Y_{t+p-i}]
		&= \sum_{j=0}^{p}\phi_{t+p-i,j} \E[X_{t}X_{t+p-i-j}]\\
		&= \phi_{t+p-i,p-i}\sigma_{X}^{2},
	\end{align*}
%		&= \E\left[X_{t}\left(\sum_{j=0}^{p}\phi_{t+p-i,j}X_{t+p-i-j}+N_{t+p-i}\right)\right] \\
for $i\geq 0$. Using the backward model of \eqref{defYt}, $\E[X_{t}Y_{t+p-i}]$ can be computed alternatively as 
\begin{align*}
	\E[X_{t}Y_{t+p-i}] &= \E\biggl[Y_{t+p-i}\biggl(\sum_{j=0}^{q}\psi_{t,j}Y_{t-j}+\widetilde{N}_{t}\biggr)\biggr]\\
	&= \sum_{j=0}^{q}\psi_{t,j}\E[Y_{t+p-i}Y_{t-j}]\\
	&= \sum_{j=\max(0,i-2p)}^{i}\psi_{t,j}\E[Y_{t+p-i}Y_{t-j}],
\end{align*}
where the last equality holds since $|t+p-i-(t-j)| \leq p$ is equivalent to $i-2p\leq j\leq i$. Thus, we obtain the following relationship for $i\geq 0$,
 \begin{equation*}
 	\sum_{j=\max(0,i-2p)}^{i}\psi_{t,j}\E[Y_{t+p-i}Y_{t-j}] = \phi_{t+p-i,p-i}\sigma_{X}^{2}.
 \end{equation*}
This can be written explicitly for $i=0$ and $i>1$ as follows, 
 \begin{equation*}
 	 \psi_{t,0}\E[Y_{t+p}Y_{t}] = 	\phi_{t+p,p}\sigma_{X}^{2}, \quad\text{ for $i=0$ }
 \end{equation*}
and for $i \geq 1$, 
 \begin{align}
 \label{pf_coro3_igeq1}
 \psi_{t,i}\E[Y_{t+p-i}Y_{t-i}]&=   \phi_{t+p-i,p-i}\sigma_{X}^{2}\nonumber\\
 &-\sum_{j=\max(0,i-2p)}^{i-1}\psi_{t,j}\E[Y_{t+p-i}Y_{t-j}]. 
 \end{align}
To further simplify the expression, we now show that $\E[Y_{t+p-i}Y_{t-i}]\ne 0$. First observe that
\begin{align*}
	\E[Y_{t}Y_{s}] &=\E\biggl[\biggl(\sum_{j=0}^{p}\phi_{t,j}X_{t-j}+N_{t}\biggr)  \left(\sum_{k=0}^{p}\phi_{s,k}X_{s-k}+N_{s}\right)  \biggr]\\
	&= \sigma_{X}^{2} \sum_{j=0}^{p}\phi_{t,j}\phi_{s,s-t+j} + \sigma_{N}^{2}\mathds{1}_{t=s},
\end{align*}
and $\E[Y_{t}Y_{s}]=0$ when $|t-s|>p$. Thus we have, for $i \geq 0$,
 \begin{align*}
 	 	\E[Y_{t+p-i}Y_{t-i}] &= \sigma_{X}^{2} \sum_{j=0}^{p}\phi_{t-i,j}\phi_{t+p-i,p+j}\nonumber\\
 	&=\sigma_{X}^{2} \phi_{t+p-i,p}\phi_{t-i,0} \neq 0, 
\end{align*}
 which follows from the fact that $\phi_{t+p-i,p+j}=0$ for $j\ge 1$. As a result, we can divide $\E[Y_{t+p-i}Y_{t-i}]$ on both sides of~\eqref{pf_coro3_igeq1} and obtain the solution
 \begin{align}
 	 \psi_{t,i}= &- \frac{\sigma_{N}^{2}\phi_{t,i-p}}{\sigma_{X}^{2} \phi_{t+p-i,p}\phi_{t-i,0}} + \frac{\phi_{t+p-i,p-i}}{\phi_{t+p-i,p}\phi_{t-i,0}}\nonumber\\
 	   &-\sum_{j=\max(0,i-2p)}^{i-1}\frac{\psi_{t,j} \sum_{k=0}^{p}\phi_{t+p-i,k}\phi_{t-j,i-j+p}}{ \phi_{t+p-i,p}\phi_{t-i,0}}
 	   \label{solution_psi}
 \end{align}
  for $i \geq 1$. Similarly, we obtain $\psi_{t,0} = 1/\phi_{t,0}$ for $i=0$. To obtain equation~\eqref{lemma2_solve}, we will need a technical lemma (postponed to be presented in Lemma~\ref{lem_noisless} below), which shows that $\Psi_{t}^{q}(\mathsf{B})$ is simply the inverse operator of $\Phi_{t}^{p}(\mathsf{B})$ if $\sigma_{N}^{2}=0$. Thus the last two terms in~\eqref{solution_psi} can be replaced by the coefficent of the inverse operator of $\Phi_{t}^{p}(\mathsf{B})$ according to~\eqref{sol_coeff}. Therefore, the coefficients of $\Psi^{q}_{t}(\mathsf{B})$ can be solved iteratively for all $i\geq 0$ using~\eqref{lemma2_solve}. 
  
  Now, we move on to the condition for the stationarity constraint. In Theorem~\ref{thm:identification}, if $\{N_{t}\}$ is an \iid process, then \eqref{var_nt} can be written as
$\Var(W_{t})= \sum_{j=1}^{\min(s,r)}\eta_{t,j}\theta_{t-j,j}\gamma_{NN}(0)$, which in turn equals to 
     $\gamma_{NN}(0)  \sum_{j=1}^{\min(q,r)}\eta_{t,j}\theta_{t-j,j}$. Since the operator $ H^{s}_{t}(\mathsf{B})$ in the backward model~\eqref{back_Hs} is uniquely determined by~\eqref{lemma2_solve}, we have $\{W_{t}\} =  \{\widetilde{N}_{t}\}$ and $\{\eta_{t,j}\} = \{\psi_{t,j}\}$. Thus, $\Var(\widetilde{N}_{t})= \sum_{j=1}^{\min(s,r)}\psi_{t,j}\theta_{t-j,j}\gamma_{NN}(0)$. The rest follows by invoking equation~\eqref{sol_coeff}.
\end{IEEEproof}
\smallskip 

It remains to show the following technical lemma. 
\begin{lemma}
	Let $\{X_{t}\}$ and $\{N_{t}\}$ be \iid processes. If $\sigma_{N}^{2}=0$ and $\sigma_{X}^{2}\neq 0$, then $\Psi_{t}^{q}(\mathsf{B})$ is the inverse operator of $\Phi_{t}^{p}(\mathsf{B})$.
	\label{lem_noisless}
\end{lemma}
\smallskip
\begin{IEEEproof}
	By replacing $Y_{t}$ in the backward model~\eqref{eq_back} with the forward model~\eqref{eq_forw}, we obtain
	\begin{equation}
\label{lem9_eq}
	X_{t} = \Psi_{t}^{q}(\mathsf{B})\Phi_{t}^{p}(\mathsf{B})X_{t} +  \Psi_{t}^{q}(\mathsf{B})N_{t} + \widetilde{N}_{t},
	\end{equation}
	where $\{X_{t}\}$ and $\{N_{t}\}$ are \iid processes. By multiplying $X_{t}$ and taking expectation to both sides of~\eqref{lem9_eq}, we obtain
\begin{align}
	\E[X_{t}^{2}] &= \psi_{t,0}\phi_{t,0}\E[X_{t}^{2}] + \E[X_{t}\widetilde{N}_{t}]\nonumber\\
	&= \psi_{t,0}\phi_{t,0}\E[X_{t}^{2}] + \E[\widetilde{N}_{t}^{2}]	,\nonumber
	\end{align}
where the last equality is obtained using the backward model. By same argument leading up to~\eqref{solution_psi}, we have $\psi_{t,0}\phi_{t,0}=1$, which implies $ \E[\widetilde{N}_{t}^{2}]=0$. It follows that $\E[\widetilde{N}_{t}\widetilde{N}_{t-k}]=0$ for any $k \in \mathbb{Z}$ by the Cauchy-Schwarz inequality. Let $\Theta_{t}^{r} \triangleq \Psi_{t}^{q}(\mathsf{B})\Phi_{t}^{p}(\mathsf{B})$. Similarly, by multiplying $X_{t-k}, k\geq 1$, and taking expectation to both sides of~\eqref{lem9_eq}, we obtain
\begin{align}
	0 = \E[X_{t}X_{t-k}] =  \theta_{t,k}\E[X_{t-k}^{2}], \nonumber
\end{align}
which implies that $ \theta_{t,k} = 0 $ for all $k\geq 1$. Then $\Theta_{t}^{r} = \Psi_{t}^{q}(\mathsf{B})\Phi_{t}^{p}(\mathsf{B}) = 1$. Therefore, $\Psi_{t}^{q}(\mathsf{B})$ is the inverse operator of $\Phi_{t}^{p}(\mathsf{B})$.
\end{IEEEproof}

\section{Proof of Corollary~\ref{coro:zero_order}}
\label{sec:coro2}

\begin{proof}
%First, we observe that $\{Y_{t}\}$ is a sequence of independent random variables. 
We solve for the operator $\Psi_{t}^{q}(\mathsf{B})$ by computing $\E[X_{t}Y_{t-j}]$ in two ways using the forward model \eqref{model_prop2} and the backward model \eqref{eq_back}, respectively. Using model~\eqref{model_prop2} we have $\E[X_{t}Y_{t-j}] = \phi(t)\sigma_{X}^{2}\mathds{1}_{j=0}$, while model \eqref{eq_back} leads to $\E[X_{t}Y_{t-j}] 
	= \psi_{t,j}\left(\phi^{2}(t-j)\sigma_{X}^{2}+\sigma_{N}^{2}\right)$.
%\begin{align}
%	\E[X_{t}Y_{t-j}] &= \E\biggl[Y_{t-j}\biggl(\sum_{k=0}^{q}\psi_{t,k}Y_{t-k}+\widetilde{N}_{t}\biggr)\biggr]\\
%	&= \sum_{k=0}^{q}\psi_{t,j}\E[Y_{t-j}Y_{t-k}]\\
%	&= \psi_{t,j}\left(\phi^{2}(t-j)\sigma_{X}^{2}+\sigma_{N}^{2}\right).
%\end{align} 
%\begin{align*}
%	\E[X_{t}Y_{t-j}] &= \sum_{k=0}^{q}\psi_{t,j}\E[Y_{t-j}Y_{t-k}]\\
%	&= \psi_{t,j}\left(\phi^{2}(t-j)\sigma_{X}^{2}+\sigma_{N}^{2}\right).
%\end{align*} 	
%By solving $\phi(t)\sigma_{X}^{2}\mathds{1}_{j=0} = \psi_{t,j}\left(\phi^{2}(t-j)\sigma_{X}^{2}+\sigma_{N}^{2}\right)$, 
Thus we obtain $\psi_{t,0} = \phi(t)/(\phi^{2}(t)+\sigma_{N}^{2}/\sigma_{X}^{2})$
%\begin{equation*}
%	\psi_{t,0} = \frac{\phi(t)}{\phi^{2}(t)+\sigma_{N}^{2}/\sigma_{X}^{2}}
%\end{equation*}
and $\psi_{t,j}=0$ for $j \geq 1$, which gives the coefficient of $Y_t$ in \eqref{back_prop1}.

Then, by substituting \eqref{model_prop2} into \eqref{back_prop1}, we obtain
 \begin{equation*}
	\widetilde{N}_{t} = \frac{\sigma_{N}^{2}/\sigma_{X}^{2}}{\phi^{2}(t)+\sigma_{N}^{2}/\sigma_{X}^{2}}X_{t} - 
\frac{\phi(t)}{\phi^{2}(t)+\sigma_{N}^{2}/\sigma_{X}^{2}}N_{t},
\end{equation*}
 which shows that $\{\widetilde{N}_{t}\}$ is a sequence of independent random variables. The rest is followed by computing the variance of $\{\widetilde{N}_{t}\}$ and write $\widetilde{N}_{t} = \sqrt{\Var(\widetilde{N}_{t})}W_{t}$, where $\{W_{t}\}$ is an \iid process with $\sigma_{W}^{2}=1$. 
\end{proof}

\section{Proof of Proposition~\ref{prop:infty}}
\label{app:prop1}
\begin{IEEEproof}
By the monotone convergence theorem and the finiteness of $\sum_{j=0}^{\infty}|\psi_{t,j}|$ and $\sup_{t}\E[|X_{t}|]$, we have
\begin{align*}
            \E\biggl[\sum_{j=0}^{\infty}|\psi_{t,j}||X_{t-j}|\biggr] &= \lim_{n \to \infty} \E\biggl[\sum_{j=0}^{n}|\psi_{t,j}||X_{t-j}|\biggr]\\
        &\leq \lim_{n \to \infty}\biggl(\sum_{j=0}^{n}|\psi_{t,j}|\biggr)\sup_{t}\E[|X_{t}|]
        < \infty,
\end{align*}
which shows that $\textstyle\sum_{j=0}^{\infty}|\psi_{t,j}||X_{t-j}|$ is finite with probability one. If $\sup_{t}\E[|X_{t}|^{2}] < \infty$ and $n>m>0$, then
\begin{align*}
 &\E\biggl[\biggl|\sum_{m<j\leq n}\psi_{t,j}X_{t-j}\biggr|^{2}\biggr] \\
 &\hspace{6em}= \sum_{m<j\leq n}\sum_{m<k \leq n}\psi_{t,j}\bar{\psi}_{t,k}\E[X_{t-j}\bar{X}_{t-k}]\\
        &\hspace{6em}\leq \biggl(\sum_{m<j\leq n}|\psi_{t,j}|\biggr)^2\sup_{t}\E[|X_{t}|^{2}]\to 0,
\end{align*}
as $n,m \to \infty$. Thus, by Cauchy criterion, the series $\textstyle\sum_{j=0}^{\infty}\psi_{t,j}X_{t-j}$ converges in mean square. Finally, let $S$ denote the mean square limit and by Fatou's lemma, 
\begin{align*}
    \E[|S-\Psi_{t}^{\infty}(\mathsf{B})X_{t}|^{2}] &= \E\biggl[\liminf_{n \to \infty} \biggl|S - \sum_{j=0}^{n}\psi_{t,j}X_{t-j} \biggr|^{2} \biggr]\\
    & \leq \liminf_{n \to \infty} \E\biggl[\biggl|S - \sum_{j=0}^{n}\psi_{t,j}X_{t-j} \biggr|^{2} \biggr]=0,
\end{align*}
which shows that the mean square limit $S$ and $\Psi_{t}^{\infty}(\mathsf{B})X_{t}$ are equal with probability one. 
\end{IEEEproof}

\section{Equivalent Definitions}
\label{app:def}
To show the equivalence of the two definitions, it suffices to prove the following direction since the other direction is trivial.
 \begin{proposition}
    \label{prop:def}
 	For two operators $\textstyle\Phi_{t}^{p}(\mathsf{B})$ and $\textstyle\Psi_{t}^{q}(\mathsf{B})$, with $p, q \in \bar{\mathbb{Z}}_{\geq 0}$, if 
 	\begin{equation}
 	 	\label{eq_poly_equal}
   \Phi_{t}^{p}(z) =  \Psi_{t}^{q}(z)
\end{equation}
holds for $z$ in some open set $\mathbb{E} \subseteq{C}$ that contains $0$, then $\Phi_{t}^{p}(\mathsf{B}) = \Psi_{t}^{q}(\mathsf{B})$.
 \end{proposition}
  \smallskip
  
\begin{IEEEproof}
We prove that $\phi_{t,j}=\psi_{t,j}$ for all $j\geq 0$ by induction. Let $z=0$ in \eqref{eq_poly_equal}, we obtain $\phi_{t,0}=\psi_{t,0}$ and $\textstyle \sum_{k=1}^{\infty}\phi_{t,k}z^{k} = \sum_{k=1}^{\infty}\psi_{t,k}z^{k}$. Assume that $\phi_{t,k}=\psi_{t,k}$ for $k \leq j$. Then, for any $z \in \mathbb{E}$, we have 
\begin{align*}
	&\textstyle \sum_{k=j+1}^{\infty}(\phi_{t,k}-\psi_{t,k})z^{k} \\
	&= z^{j+1}\sum_{k=0}^{\infty}(\phi_{t,k+j+1}-\psi_{t,k+j+1})z^{k} = 0. 
\end{align*}
It follows that, we have $\sum_{k=0}^{\infty}(\phi_{t,k+j+1}-\psi_{t,k+j+1})z^{k} = 0 $ for any $z \in \mathbb{E} \setminus 0$. Finally, taking $\lim_{z \to 0}$ on both sides of the last equality yields $\phi_{t,j+1} = \psi_{t,j+1}$. Therefore, we have proved that $\phi_{t,j}=\psi_{t,j}$ for all $j \geq 0$ as claimed.
\end{IEEEproof}

\section{Technical Lemmas}
\label{app:determinant}
\smallskip
\begin{lemma}
\label{lem:cov}
For a Gaussian process $\{Y_t\}$ and a lag operator $\Phi_t^p(\mathsf{B})$, $p \in \bar{\mathbb{Z}}_{\geq 0}$, we have that $\{Y_t\}$ is independent of $\{\Phi_t^p(\mathsf{B})Y_t\}$ only if the covariance function of $\{\Phi_t^p(\mathsf{B})Y_t\}$ is zero. 
\end{lemma}
% $\Cov(\Phi_{t_{1}}^{p}(\mathsf{B})Y_{t_{1}},\Phi_{t_{2}}^{p}(\mathsf{B})Y_{t_{2}})=0$
\smallskip

\begin{IEEEproof} The claim is trivial when $\Phi_t^p(\mathsf{B})=0$. Suppose that $\{Y_t\}$ and $\{\Phi_{t}^{p}(\mathsf{B})Y_{t}\}$ are independent, which implies that 
\begin{align}
	\Cov(Y_{t_1}, \Phi_{t_{2}}^{p}(\mathsf{B})Y_{t_{2}}) = \Phi_{t_{2}}^{p}(\mathsf{B})  \gamma_{YY}(t_{1},t_{2}) = 0. \label{lem3eq:cov}
\end{align} 
By applying $\Phi_{t_{1}}^{p}(\mathsf{B})$ to~\eqref{lem3eq:cov}, we have that 
\begin{align*}
	\Phi_{t_{1}}^{p}(\mathsf{B}) \Phi_{t_{2}}^{p}(\mathsf{B})  \gamma_{YY}(t_{1},t_{2}) = 0,
\end{align*}
where the left-hand side is simply the covariance function of $\{\Phi_t^p(\mathsf{B})Y_t\}$. 
\end{IEEEproof}

The following technical lemma contains a list of basic properties of  matrix norms (see proofs in~\cite{horn2012matrix}).

\smallskip
\begin{lemma}
\label{lem:determinant}
For $A \in \mathbb{C}^{n \times n}$ and $x \in \mathbb{C}^{n \times 1}$, we have 
\begin{enumerate}[(1)]
	\item $||Ax||_{p} \leq ||A||_{p} ||x||_{p}$. \label{3.1}
	\item $||A_{1} A_{2}\ldots A_{k}||_{p}\leq ||A_{1}||_{p}||A_{2}||_{p} \ldots ||A_{k}||_{p}$, where $A_{1}, A_{2},\ldots, A_{k} \in \mathbb{C}^{n \times n}$.\label{3.2}
	\item For any two matrix norms, there exists a constant $0<C_{\alpha\beta}<\infty$ such that $||A||_{\alpha} \leq C_{\alpha\beta} ||A||_{\beta}$ for any matrix $A\in \mathbb{C}^{n \times n}$.\label{3.3}
	\item For any $\delta>0$, there exists a matrix norm $||\cdot||_{*}$ such that $0 \leq ||A||_{*}-\rho(A)\leq \delta$.\label{3.4}
%	\item $||A||_{1} \leq \sqrt{n} ||A||_{2}$.
%	\item $||A||_{2} = \max_{1\leq j \leq p} |\lambda_{j}|$, where $\{\lambda_{1},\ldots,\lambda_{n}\}$ are the eigenvalues of $A$.
%	\item $||A_{1}A_{2}\ldots A_{k} ||_{p} \leq ||A_{1}||_{p} ||A_{2}||_{p} \dots ||A_{k}||_{p}$, where $A_i \in \mathbb{C}^{n \times n}$ for $1\le i\le k$.
\end{enumerate}
\end{lemma}
\smallskip

The last two lemmas focus on a particular form of matrix called the \emph{companion matrix}~\cite{key2014note}. For a product of companion matrices, the following lemma provides a condition for its spectral radius to be bounded by an exponentially decreasing sequence.
\smallskip  
\begin{lemma}[\cite{key2014note}]
\label{lem:companion}
Let $A_{i} \in \mathbb{R}^{n \times n}, 1 \leq i \leq k$, be companion matrices of the form
\begin{equation}
\label{companion}
  A_{i} = 
	\left[
\begin{array}{c|cc}
  \begin{matrix}
  0   \\
  \vdots  \\
  0 
  \end{matrix} &  \mbox{\normalfont\Large\bfseries $I_{n-1}$}\\ 
  \hline
\begin{matrix}
  -a_{i,n}
\end{matrix} &
  \begin{matrix}
   -a_{i,n-1} & \ldots & -a_{i,1}
  \end{matrix} \\
\end{array}
\right],
\end{equation}
where $I_{n}$ denotes the $n \times n$ identity matrix. If $a_{i,0} \triangleq 1 > a_{i,1} > \ldots > a_{i,n-1} > a_{i,n} \geq 0$ 
for each $i$, then there exists 
\begin{equation*}
	\varepsilon = \max_{1 \leq i \leq k, 1 \leq j \leq n}\frac{a_{i,j}}{a_{i,j-1}}<1
\end{equation*}
such that $\rho(A_{k}A_{k-1}\ldots A_{1}) \leq \varepsilon^{k}<1$.
\end{lemma}
\smallskip

Finally, we establish the following lemma on a product of companion matrices inspired by~\cite{bauer1993stability}.
\smallskip
\begin{lemma}
\label{companion2}
Let $A_{i} \in \mathbb{R}^{n \times n}, 1 \leq i \leq n$, be companion matrices of the form in~\eqref{companion}. If
\begin{equation}
	 \label{cond_lm5}
	 0<\sum_{j=1}^{n}|a_{i,j}|<1
\end{equation}
for $1 \leq i \leq n$, then there exists some $0<\varepsilon<1$ such that $||A_{n}A_{n-1}\ldots A_{1}||_{\infty}\leq \varepsilon$. Moreover, for companion matrices $A_{i}$ with $1 \leq i \leq Nn$ and $N \geq 1$, if~\eqref{cond_lm5} hold for $1 \leq i \leq Nn$, then
\begin{equation*}
	 ||A_{Nn}A_{Nn-1}\ldots A_{1}||_{\infty}\leq \varepsilon^{N}
\end{equation*}
for some $0<\varepsilon<1$.
\end{lemma}
\smallskip
\begin{IEEEproof}
Let $T_{i}\triangleq A_{i}A_{i-1}\ldots A_{1}$, $1\leq i\leq n$, and let $t_{j}^{(i)}, 1 \leq j \leq n,$ denote the $j^{th}$ row of $T_{i}$. We first claim that
\begin{align}
	\label{induct}
	||t_{j}^{(i)}||_{1} 
\begin{cases}
	=1, &1 \leq j \leq n-i,\\
	<1, &n-i+1 \leq j \leq n,
\end{cases}		
\end{align}
for $1 \leq i \leq n-1$, and
\begin{equation}
\label{induct2}
	||t_{j}^{(n)}||_{1}<1, \quad 1 \leq j \leq n,
\end{equation}
 which implies that $||T_{n}||_{\infty}\leq \varepsilon <1$ for some $0<\varepsilon<1$. 

Now, we prove this claim by induction. For $i=1$, the statement  follows directly from the assumption that $ 0<\sum_{j=1}^{n}|a_{1,j}|<1$. For any $1 \leq i \leq n-1$, if \eqref{induct} holds, then following from the structure of $A_{i}$ in \eqref{companion}, we obtain
\begin{align*}
||t^{(i+1)}_{j}||_{1}=
\begin{cases}
	1, & 1 \leq j \leq n-i-1, \\
	||t^{(i)}_{j+1}||_{1}<1, &  n-i \leq j \leq n-1,
\end{cases}
\end{align*}
and 
\begin{align*}
	||t^{(i+1)}_{n}||_{1}
	&\leq \sum_{k=1}^{n} |a_{i+1,k}|||t_{n-k+1}^{(i)}||_{1}
	\overset{(a)}{\leq} \sum_{k=1}^{n} |a_{i+1,k}|<1,
	\end{align*}
where (a) is due to $||t_{n-k+1}^{(i)}||_{1}\leq 1$ from~\eqref{induct} and~\eqref{induct2}. Thus, we have proved the first part of the lemma by induction.

For the second part, by Lemma~\ref{lem:determinant}.\ref{3.3} combined with the first part, we find that
\begin{align*}
	 &||A_{Nn}A_{Nn-1}\ldots A_{1}||_{\infty} \nonumber\\
	  &\hspace{5em}\leq \prod_{k=1}^{N}||A_{kn}A_{kn-1}\ldots A_{kn-n+1} ||_{\infty}\leq \varepsilon^{N}	  
\end{align*}
for some $0<\varepsilon<1$, as claimed.
\end{IEEEproof}

\section{Proof of Lemma~\ref{lem:inverse}}
\label{app:lemma1}

According to~\cite[equation~(4.10)]{abdrabbo1967prediction}), $\phi_{t,0}\neq 0$ is a necessary condition for $\Phi_{t}^{p}(\mathsf{B})$ to be invertible and the coefficients of the inverse operator $\Theta_{t}^{r}(\mathsf{B})$ can be solved iteratively by
\begin{numcases}{\theta_{t,i} = }
1/\phi_{t,0} , & $i = 0$, \nonumber \\
-(1/\phi_{t-i,0})\cdot\sum_{j = 1}^{i}\theta_{t,i-j}\phi_{t-i+j,j} , & $1 \leq  i \leq p-1$,\nonumber \\
-(1/\phi_{t-i,0})\cdot\sum_{j = 1}^{p}\theta_{t,i-j}\phi_{t-i+j,j}, & $i \geq p$. \label{eq_solve}
\end{numcases}
For a fixed $t$, we take $\{\theta_{t,i}$, $0 \leq i \leq p-1\}$ as the initial value, then \eqref{eq_solve} is a homogeneous linear difference equation, which can be represented in a multi-dimensional form
\begin{equation}
\label{lde_mul}
    \boldsymbol{x}_{t,n} = 
    \begin{cases}
    A_{t,n}\boldsymbol{x}_{t,n-1},& n \geq 1,\\
    \boldsymbol{x}_{t,0}, & n=0,
    \end{cases}
\end{equation}
where $\boldsymbol{x}_{t,n} = [\theta_{t,n},\ldots,\theta_{t,n+p-1}]^{T}$ and
\begin{equation}
\label{mat_atn}
  A_{t,n} =  
	\left[
\begin{array}{c|cc}
  \begin{matrix}
  0   \\
  \vdots  \\
  0 
  \end{matrix} &  \mbox{\normalfont\Large\bfseries $I_{p-1}$}\\ 
  \hline
\begin{matrix}
 -a_{t,n,p}
 \end{matrix} &
  \begin{matrix}
   -a_{t,n,p-1} & \ldots &  -a_{t,n,1}
  \end{matrix} \\
\end{array}
\right]
\end{equation}
with $a_{t,n,j} \triangleq  (\phi_{t-(n+p-1)+j,j})/(\phi_{t-(n+p-1),0})$ for $1 \leq j \leq p$. By the Leibniz formula of determinant~\cite{horn2012matrix}, we obtain
\begin{equation*}
    |A_{t,n}| = (-1)^{p+1} a_{t,n,p}. 
\end{equation*}
Given the initial value $\boldsymbol{x}_{t,0}$, the solution of equation \eqref{lde_mul} is given by
\begin{equation*}
    \boldsymbol{x}_{t,n} = A_{t,n}A_{t,n-1}\ldots A_{t,1}\boldsymbol{x}_{t,0} \triangleq T_{t,n}\boldsymbol{x}_{t,0},
\end{equation*}
where $\boldsymbol{x}_{t,0}$ is constantly non-zero due to $\phi_{t,0} \neq 0$.

\smallskip
\begin{IEEEproof}[Proof of Lemma~\ref{lem:inverse}]
(I) We start with the sufficient conditions. Recall the condition~\eqref{lemma_condi_a} $|\phi_{t,0}|> \sum_{j=1}^{p}|\phi_{t+j,j}|>0$ in Lemma~\ref{lem:inverse}. Note that $\phi_{t,0} \neq 0$, for all $t$, follows directly from this sufficient condition. We will show that if~\eqref{lemma_condi_a} holds, then an inverse operator $\Theta_{t}^{r}(\mathsf{B})$ exists. It suffices to prove that the coefficients of $\Theta_{t}^{r}(\mathsf{B})$ are absolutely summable. This is trivial when $r$ is finite. The remainder of the proof is thus devoted to the case when $r=\infty$.

First, since condition \eqref{lemma_condi_a} implies that $\sum_{j=1}^{p}|a_{t,n,j}|<1$, 
for matrix $A_{t,n}$ in \eqref{mat_atn}, we have $||A_{t,n}||_{\infty}\leq 1$ for all $n\geq 1$. It follows that $||T_{t,n}||_{\infty} = ||A_{t,n}T_{t,n-1}||_{\infty}\leq  ||A_{t,n}||_{\infty} ||T_{t,n-1}||_{\infty}  \leq ||T_{t,n-1}||_{\infty}$, where the first inequality is from Lemma~\ref{lem:determinant}.\ref{3.1}. We thus observe that $||T_{t,n}||_{\infty}$, for $n\geq 1$, is a non-increasing sequence in $n$. 

Note that the sequence of companion matrices $A_{t,n}$ satisfies condition~\eqref{cond_lm5} in Lemma \ref{companion2}, thus there exists $0<\varepsilon<1$ such that $||T_{t,jp+k}||_{\infty} \leq ||T_{t,jp}||_{\infty}\leq \varepsilon^{j}$, 
for any $j \geq 0$ and $1 \leq k \leq p$.

Now we show that the coefficients of $\Theta_{t}^{\infty}(\mathsf{B})$ are absolutely summable. Note that $\sum_{i=0}^{\infty}|\theta_{t,i}| \leq ||\boldsymbol{x}_{t,0}||_{1} + \frac{1}{p} \sum_{i=1}^{\infty} ||\boldsymbol{x}_{t,i}||_{1}$ due to the additional non-negative terms. We can upper bound $\frac{1}{p} \sum_{i=1}^{\infty} ||\boldsymbol{x}_{t,i}||_{1}$ as follows, 
\begin{align}
 \frac{1}{p} \sum_{i=1}^{\infty} ||\boldsymbol{x}_{t,i}||_{1}
   &\leq   \frac{||\boldsymbol{x}_{t,0}||_{1}}{p}\sum_{i=1}^{\infty}||T_{t,i}||_{1}\label{eq_b}\\
  &\leq  \frac{||\boldsymbol{x}_{t,0}||_{1}C}{p}\sum_{i=1}^{\infty}||T_{t,i}||_{\infty}\label{eq_c}\\
        &=   \frac{||\boldsymbol{x}_{t,0}||_{1}C}{p}\sum_{j=0}^{\infty}\sum_{k=1}^{p}||T_{t,jp+k}||_{\infty}\nonumber\\
        &\leq  ||\boldsymbol{x}_{t,0}||_{1}C\sum_{j=0}^{\infty}\varepsilon^{j}<\infty\nonumber,
\end{align}
where $\eqref{eq_b}$ and $\eqref{eq_c}$ use Lemma~\ref{lem:determinant}.\ref{3.1} and Lemma~\ref{lem:determinant}.\ref{3.3}, respectively. Therefore we have $\sum_{i=0}^{\infty}|\theta_{t,i}| <\infty$. This completes the proof of the sufficient condition~\eqref{lemma_condi_a} in Lemma~\ref{lem:inverse}.

Recall the second sufficient condition in~\eqref{lemma_condi_b} $\phi_{t,0} > \phi_{t+1,1} > \ldots > \phi_{t+p,p} \geq 0$. Note again that $\phi_{t,0} \neq 0$, for all $t$, follows directly from this sufficient condition. Similar to the first sufficient condition, we focus on the case when $r=\infty$, and show that if~\eqref{lemma_condi_b} holds, then the coefficients of $\Theta_{t}^{\infty}(\mathsf{B})$ are absolutely summable. Since~\eqref{lemma_condi_b} implies that 
\begin{equation*}
	a_{t,n,0} = 1 > a_{t,n,1} > \ldots > a_{t,n,p-1} > a_{t,n,p} \geq 0, 
\end{equation*} 
it follows from Lemma~\ref{lem:companion} that there exists $0<\varepsilon<1$ such that $\rho(T_{t,n}) \leq \varepsilon^{n}$. 
Then by Lemma~\ref{lem:determinant}.\ref{3.4}, there exists a matrix norm $||\cdot||_{(n)}$ such that $||T_{t,n}||_{(n)} \leq \rho(T_{t,n})+2^{-n}$ for each $n \geq 1$. Hence there exists $0<C_{n}<\infty$ and $0 < \varepsilon < 1$ such that
\begin{align*}
||T_{t,n}||_{1} &\leq	C_{n} ||T_{t,n}||_{(n)}\\
 &\leq C_{n} \left( \rho(T_{t,n})+2^{-n} \right)
 \leq C_{n} \left( \varepsilon^{n}+2^{-n} \right),
\end{align*} 
where the first inequality follows from Lemma~\ref{lem:determinant}.\ref{3.3}. Using equation~\eqref{eq_b} again, we find that the coefficients of $\Psi^{\infty}_{t}(\mathsf{B})$ are absolute summable since 
\begin{align*}
   \frac{||\boldsymbol{x}_{t,0}||_{1}}{p}\sum_{i=1}^{\infty}||T_{t,i}||_{1}
        &\leq  \frac{||\boldsymbol{x}_{t,0}||_{1}}{p}\sum_{i=1}^{\infty}C_{i}(\varepsilon^{i}+2^{-i})\\   
        & \stackrel{(d)}{\leq} \frac{||\boldsymbol{x}_{t,0}||_{1}C}{p}\sum_{i=1}^{\infty}(\varepsilon^{i}+2^{-i}) < \infty,
\end{align*}
where $(d)$ is due to $ 0<C \triangleq \max_{i \geq 1}C_{i}<\infty$. Putting together the pieces yields the two sufficient conditions.  
   
\smallskip
(II) Now we move on to the necessary condition. Assume that $\Theta_{t}^{q}(\mathsf{B})$ exists for finite $q$, then $\phi_{t,0}\ne 0$~\cite{abdrabbo1967prediction}. Recall $\boldsymbol{x}_{t,n} = [\theta_{t,n},\ldots,\theta_{t,n+p-1}]^{T}$ and note that we have $\boldsymbol{x}_{t,q+1} = 0$ due to the finiteness of $q$. This leads to $\boldsymbol{x}_{t,q+1} = T_{t,q+1}\boldsymbol{x}_{t,0} = 0$, where the only solution of this homogeneous linear system is zero if $T_{t,q+1}$ is nonsingular\cite{horn2012matrix}. However, 
since $\boldsymbol{x}_{t,0}$ is non-zero, we must have that $T_{t,q+1}$ is singular, i.e.,
\begin{equation*}
    |T_{t,q+1}|= \prod_{i=1}^{q+1}|A_{t,i}| = 0.
\end{equation*}
This implies that $\prod_{i=0}^{q}\phi_{t-i,p}=0$. Combined with $\phi_{t,0}\ne 0$, we have shown the necessary 
condition~\eqref{nece} in Lemma~\ref{lem:inverse}, as claimed.

\end{IEEEproof}

\bibliographystyle{IEEEtran}
\balance

\bibliographystyle{IEEEtran}
\balance

\bibliography{ref}\bibliographystyle{IEEEtran}
\balance

%\bibliography{ref}

%\appendix

%\section{Proof of Lemma~\ref{lemma:recursive}}
%\label{sec:proof_recursive}

\end{document}